\documentclass[reprint,amsmath,amssymb,aip]{revtex4-2}
\usepackage{graphicx}
\usepackage{dcolumn}
\usepackage{bm}
\begin{document}

\preprint{}

\title{Acceleration Theorem for Low-Dimensional Electron Systems with
  Off-Diagonal Effective Mass Components}

\author{Nobuya Mori}
\email{nobuya.mori@eei.eng.osaka-u.ac.jp}
\author{Hajime Tanaka}
\author{Jo Okada}
\affiliation{Division of Electrical, Electronic and Infocommunications
  Engineering, Osaka University, Suita, Osaka, 565–0871, Japan}

\date{\today}

\begin{abstract}
  The motion of electrons under homogeneously applied electric fields
  in low-dimensional systems with non-zero off-diagonal effective mass
  (ODEM) is studied.  The equation describing the time evolution of a
  probability coefficient of finding an electron in a subband is
  derived using the Krieger-Iafrate theory in the effective mass
  approximation.  It is shown that an electron can change subbands
  during free flight due to the ODEM-induced inter-subband
  transitions.  By introducing an effective dispersion defined as a
  weighted average of the subband dispersions, it is also shown that
  the initial acceleration of an electron effectively follows the bulk
  dispersion relation.  The results obtained suggest that the
  transport properties of the quantized systems when many subbands are
  occupied in the weak confinement limit approach the values one would
  find without considering the quantization.
\end{abstract}

%\keywords{Suggested keywords}

\maketitle

\section{Introduction}

Silicon belongs to the cubic crystal system and has isotropic carrier
mobility.\cite{Nye1985} However, in a confined structure such as
silicon-on-insulator (SOI) or nanosheet, the mobility depends on the
surface and channel crystallographic
orientations.\cite{Stern1967,Ando1982,Esseni2009a,Esseni2011} By
utilizing this dependence to enhance the channel mobility, we can
improve device
performance.\cite{Takagi1994,Laux2004,Low2004,Irie2004,Tsutsui2006,Chen2009,Mochizuki2024}
To design such devices, predictive physical models of the carrier
transport properties for different crystallographic orientations are
required.\cite{Silvestri2010}

\begin{figure}[b]%[htbp]
  \centering
  \includegraphics[scale=0.3]{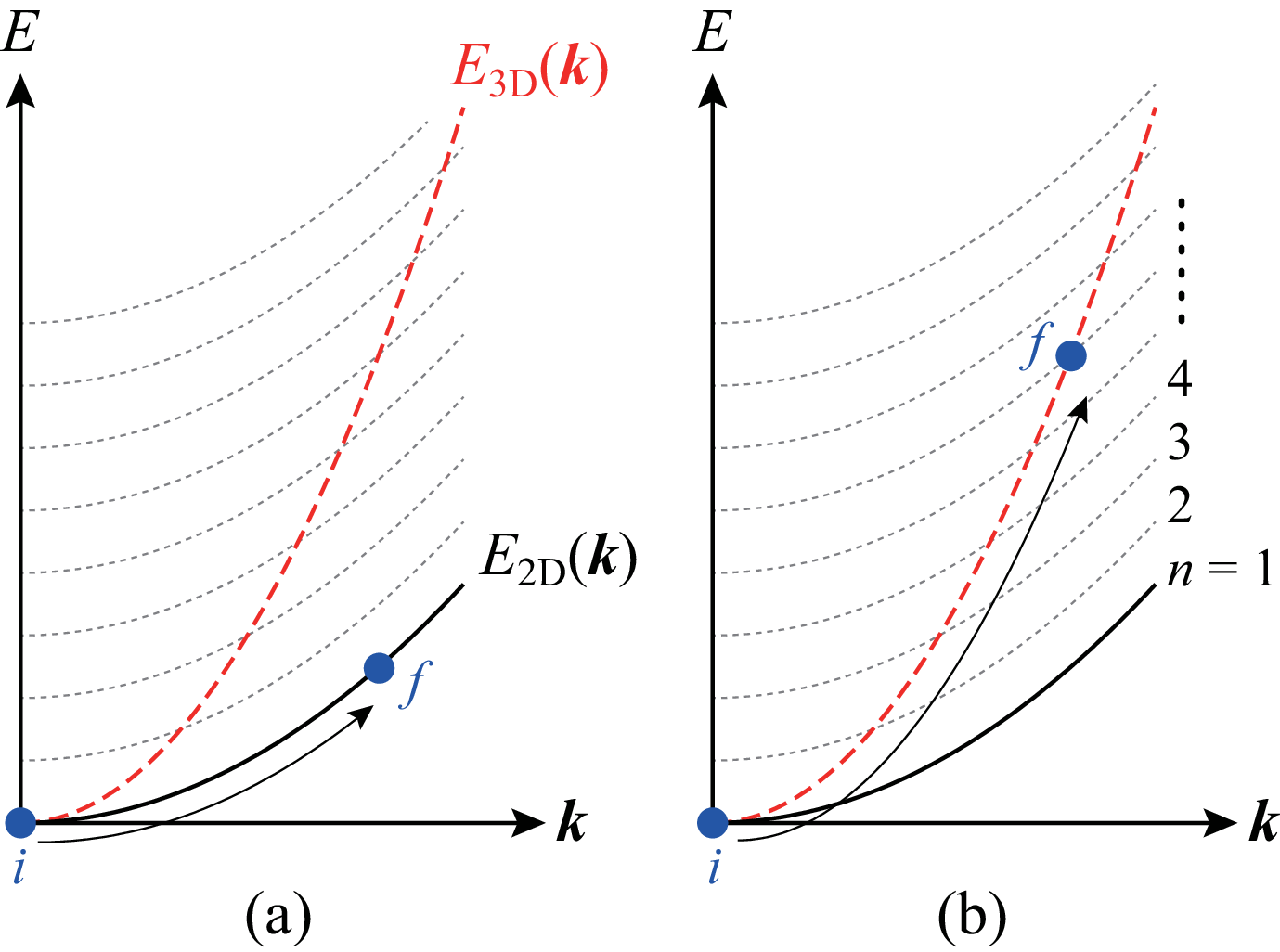}
  \caption{An electron is accelerated from state $i$ to state $f$ by
    an electric field.  The black solid line represents
    $E_\mathrm{2D}(\bm{k})$ and the red dashed line
    $E_\mathrm{3D}(\bm{k})$. The gray dotted lines correspond to the
    upper subband dispersions.  (a) If we neglect the inter-subband
    transitions, an electron remains in a subband during free flight.
    This implies that, in the weak confinement limit, the 2DEG
    transport properties do not approach the corresponding bulk
    values.  (b) The off-diagonal effective mass-induced inter-subband
    transitions promote an electron to upper subbands, so that the
    acceleration of the electron effectively follows the bulk
    dispersion in an initial period of time of the
    acceleration. \label{fig:odem}}
\end{figure}

Stern and Howard \cite{Stern1967} developed a theory within the
effective-mass approximation to calculate energy levels with arbitrary
surface and channel orientations for a two-dimensional electron gas
(2DEG) in a metal-oxide-semiconductor (MOS) structure.  For a 2DEG
moving in the $(x, y)$ (or $(x_1, x_2)$) plane, they showed that the
subband level $E_n$ associated with the quantized $z$-motion does not
depend on the in-plane wavevector $\bm{k}$ ($ = (k_x, k_y)$) when a
vanishing wavefunction is assumed at the interface. The in-plane
dispersion measured from $E_n$ is then given by the equation
\begin{align}
  E_{\mathrm{2D}}(\bm{k})
  &
  = 
  \mbox{$\frac{1}{2}$}\hbar^2 \biggl[
  \left(w_{11} - \frac{w_{13}^2}{w_{33}} \right) k_x^2 
  + 
  \left(w_{22} - \frac{w_{23}^2}{w_{33}} \right) k_y^2
  \nonumber
  \\
  &
  \qquad
  \qquad
  + 
  2 \left(w_{12} - \frac{w_{13} w_{23}}{w_{33}} \right) k_x k_y 
  \biggr],
\end{align}
where $w_{ij}$ is the reciprocal effective-mass tensor.  
% This $E_{\mathrm{2D}}(\bm{k})$
It can be written as
$E_{\mathrm{2D}}(\bm{k}) = E_{\mathrm{3D}}(\bm{k}) - \frac{1}{2}
\hbar^2 w_{33} (\bm{\gamma}\cdot\bm{k})^2$, where
$E_{\mathrm{3D}}(\bm{k}) =\frac{1}{2} \hbar^2 (w_{11} k_x^2 + 2 w_{12}
k_x k_y + w_{22} k_y^2)$ is the in-plane dispersion in the bulk and
$\bm{\gamma} = (w_{13} / w_{33}, w_{23} / w_{33})$.  The 2DEG in-plane
kinetic energy $E_\mathrm{2D}(\bm{k})$ therefore differs from the bulk
dispersion $E_{\mathrm{3D}}(\bm{k})$ by
$\frac{1}{2} \hbar^2 w_{33} (\mbox{$\bm{\gamma}\cdot\bm{k}$})^2$.  One
might think that this leads to the following conclusion: the in-plane
2DEG mobility when many subbands are occupied does not approach the
value one would find without considering the surface quantization, see
Fig.~\ref{fig:odem}(a).  In the present study, we argue that this is
not the case when the off-diagonal effective mass (ODEM) induced
inter-subband transition \cite{Esseni2009b} is considered, see
Fig.~\ref{fig:odem}(b).  In a semi-classical picture, if there is an
off-diagonal component of the effective mass, electrons that are
forced by the in-plane electric field will also be accelerated in the
out-of-plane direction. This acceleration leads to inter-subband
transitions in a quantum mechanical description.

In an applied electric field, an electron in a band is accelerated in
$k$-space according to the acceleration
theorem,\cite{Jones1934,Houston1940,Kittel1987} while it remains in
the original band except for extreme cases, such as very high fields
\cite{Zener1934,Keldysh1958,Kane1959} and very short time
intervals.\cite{Adams1956,Duque-Gomez2012,Fang2014,Chang2014} In the
latter case, the time scale is determined by the band gap and is on
the order of femtoseconds in typical semiconductor systems, resulting
in no significant effect observed.\cite{Chang2014}  However, for 2DEG
with non-zero ODEM, electrons can change subbands even during
low-field free flight.\cite{Esseni2009b}  We have analyzed the 2DEG and 1DEG
motion considering this inter-subband transition with the Krieger and
Iafrate theory \cite{Krieger1986} and find that the average energy
gain is the same as in the bulk in an initial period of time of the
acceleration.

In this paper we study the electron motion in 1D and 2D systems with
non-zero ODEM in the presence of a homogeneous electric field.  In
particular, we aim to clarify the role of the ODEM-induced
inter-subband transition on the acceleration of the electrons.  We
proceed as follows.  First, the 2DEG states in the absence of an
applied electric field are determined.  The method follows the theory
of Stern and Howard,\cite{Stern1967} but for the sake of completeness
and to introduce notations, it is described first.  Then, we study the
electron acceleration under an applied electric field using the
Krieger and Iafrate equation, and show that the average energy gain is
equal to that of the bulk in an initial period of time of the
acceleration.  We then develop the same theory for 1DEG using the
theory of Bescond \textit{et al.\@} \cite{Bescond2007} for the 1DEG
states.  After describing the theory for 1DEG and 2DEG systems,
although it is applicable to general semiconductor systems, we give
some numerical examples for the case of Si, including the
effective dispersions and the drift velocities.

The organization of this paper is as follows. In
Sec.~\ref{sec:theory}.A, we describe the 2D electronic states in the
absence of an applied electric field.  We then derive an equation for
the time evolution of a probability coefficient of each subband in the
presence of a homogeneous electric field.  We show that the average
energy gain is identical to the bulk case in the short time limit.
Sec.~\ref{sec:theory}.B is devoted to the theory for 1DEG.  Some
numerical results and discussion are then given in
Sec.~\ref{sec:results}.  A summary and conclusion are given in
Sec.~\ref{sec:conclusion}.

\section{Theory\label{sec:theory}}

\subsection{Acceleration of 2DEG}

We consider 2D electrons moving freely in the $(x, y)$ plane and
confined along the $z$-direction by a potential $V(z)$ with no
electric field applied in the $(x, y)$ plane.  The effective-mass
equation is given by
\begin{align}
  \biggl[
  \frac{1}{2}
  \sum_{ij}
  w_{ij} p_i p_j
  +
  V(z)
  \biggr]
  \phi(x, y, z)
  =
  E
  \phi(x, y, z),
  \label{eq:2DEG:waveeq}
\end{align}
where $p_j = -i\hbar (\partial / \partial x_j)$,
$(x_1, x_2, x_3) = (x, y, z)$, and $w_{ij}$ ($=w_{ji}$) is the
reciprocal effective-mass tensor.  We assume that the potential $V(z)$
is high enough at the boundary plane so that the wavefunction
$\phi(x, y, z)$ vanishes there, which ensures that the subband level
does not depend on the in-plane motion.  Following the theory of
Ref.~\onlinecite{Stern1967}, we separate the 3D effective-mass
equation of Eq.~(\ref{eq:2DEG:waveeq}) into the 2D part describing
the free in-plane motion and the 1D part the quantized $z$-motion. The
latter is determined by the following effective-mass equation:
\begin{align}
  H_z \zeta_n(z)
  =
  E_n
  \zeta_n(z),
\end{align}
with
\begin{align}
  H_z
  =
  -
  \frac{\hbar^2}{2m_3}
  \frac{\mathrm{d}^2}{\mathrm{d}z^2}
  +
  V(z),
\end{align}
where $m_3 = w_{33}^{-1}$, $n$ is a subband index
($n = 1, 2, 3, \ldots$), $E_n$ is the subband level, and $\zeta_n(z)$
is the 1D wavefunction.  Using $E_n$ and $\zeta_n(z)$, we have the
energy level $E$ and the 3D wavefunction $\phi(x, y, z)$ of
Eq.~(\ref{eq:2DEG:waveeq}) as follows:
\begin{align}
  E_{n\bm{k}}
  &
  = 
  E_n + E_{\mathrm{2D}}(\bm{k}),
  \label{eq:2DEG:En}
  \\
  \phi_{n\bm{k}}(\bm{x}, z)
  &
  =
  \frac{1}{\sqrt{S}}
  \mathrm{e}^{i\bm{k}\cdot\bm{x}}
  u_{n\bm{k}}(z),
  \label{eq:2DEG:phink}
\end{align}
with
\begin{align}
  u_{n\bm{k}}(z)
  &
  =
  \mathrm{e}^{-i\bm{\gamma} \cdot \bm{k}\,z}
  \zeta_n(z).
  \label{eq:2DEG:un}
\end{align}
Here, $\bm{x} = (x, y)$, $\bm{k}$ ($= (k_x, k_y)$) is an in-plane
wavevector, $S$ is the area of the system, and
\begin{align}
  \bm{\gamma} = ( w_{13}/w_{33}, w_{23}/w_{33}).
\end{align}
The in-plane dispersion of 2DEG, $E_{\mathrm{2D}}(\bm{k})$, is given
by
\begin{align}
  E_{\mathrm{2D}}(\bm{k})
  &
  =
  E_{\mathrm{3D}}({\bm{k}})
  -
  \mbox{$\frac{1}{2}$}\hbar^2 w_{33} (\bm{\gamma} \cdot \bm{k})^2,
  \label{eq:2DEG:Ek}
\end{align}
where $E_{\mathrm{3D}}(\bm{k})$
($=\frac{1}{2} \hbar^2 (w_{11} k_1^2 + 2 w_{12} k_1 k_2 + w_{22}
k_2^2)$) is the in-plane dispersion in the bulk. The 2DEG in-plane
energy $E_{\mathrm{2D}}(\bm{k})$ is different from
$E_{\mathrm{3D}}(\bm{k})$ by
$\frac{1}{2} \hbar^2 w_{33} (\bm{\gamma}\cdot\bm{k})^2$.  Note that,
in the case where $w_{33} > 0$ as in the Si conduction band,
$\frac{1}{2} \hbar^2 w_{33} (\bm{\gamma}\cdot\bm{k})^2 > 0$
($\bm{k} \neq 0$) if $w_{13} \neq 0$ or $w_{23} \neq 0$.

We next consider the electron motion in the presence of a homogeneous
in-plane electric field $\bm{F} = (F_x, F_y)$ ($=(F_1, F_2$))
following the theory of Ref.~\onlinecite{Krieger1986}.  Using a vector
potential to describe the electric field, the time-dependent
effective-mass equation can be written as
\begin{align}
  i\hbar \frac{\partial}{\partial t} \psi(\bm{x}, z, t)
  =
  H' \psi(\bm{x}, z, t),
  \label{eq:2DEG:tdepeq}
\end{align}
where
\begin{align}
  H'
  =
  \frac{1}{2}
  \sum_{ij}
  w_{ij}
  (p_i + eA_i)
  (p_j + eA_j)
  +
  V(z),
\end{align}
and $(A_1, A_2, A_3) = (-F_1 t, -F_2 t, 0)$.  Note that the elementary
charge $e$ is defined as a positive value in this paper.  As a basis
for the expansion of $\psi(\bm{x}, z, t)$, it is convenient to
introduce the instantaneous solutions, $E'$ and $\phi'(\bm{x}, z, t)$,
of the Hamiltonian $H'$, i.e.,
\begin{align}
  H' \phi'(\bm{x}, z, t)
  =
  E' \phi'(\bm{x}, z, t).
\end{align}
Using the solutions, $E_{n\bm{k}}$ and
$\phi_{n\bm{k}}(\bm{x}, z)$, of Eq.~(\ref{eq:2DEG:waveeq}), the
instantaneous solutions are written as follows:
\begin{align}
  E'_{n\bm{k}}
  &
  = 
  E_{n\bm{k}},
  \\
  \phi'_{n\bm{k}}(\bm{x}, z, t)
  &
  =
  \mathrm{e}^{ie\bm{F}\cdot\bm{x}\,t/\hbar}
  \phi_{n\bm{k}}(\bm{x}, z).
\end{align}
We apply periodic boundary conditions on
$\phi'_{n\bm{k}}(\bm{x}, z, t)$ in the $(x, y)$-plane, which leads to
the time-dependent $\bm{k}$:\cite{Krieger1986}
\begin{align}
  \bm{k}(t) = \bm{k}(0) - \frac{e\bm{F}t}{\hbar}.
  \label{eq:2DEG:accel}
\end{align}
Expanding $\psi(\bm{x}, z, t)$ by $\phi'_{\nu}(\bm{x}, z, t)$
($\nu = (n, \bm{k})$):
\begin{align}
  \psi(\bm{x}, z, t)
  =
  \sum_{\nu}
  a_{n}(t)
  \exp\left[
  -\frac{i}{\hbar} \int_0^t \! E_{n\bm{k}(t')}
  \mathrm{d}t'
  \right]
  \phi'_{\nu}(\bm{x}, z, t),
  \label{eq:2DEG:psi}
\end{align}
and substituting this into Eq.~(\ref{eq:2DEG:tdepeq}), we have the
following Krieger and Iafrate (KI) equation \cite{Krieger1986} for the
probability coefficient $a_n(t)$:
\begin{align}
  \frac{\mathrm{d}a_n(t)}{\mathrm{d}t}
  &
  =
  -\frac{1}{i\hbar}
  \sum_{n'}
  e\bm{F}
  \cdot
  \bm{X}_{nn'}(\bm{k}(t))
  \nonumber 
  \\
  &
  \times
  \exp\left[
  \frac{i}{\hbar}
  \int_0^t \!
  \{E_{n\bm{k}(t')}-E_{n'\bm{k}(t')}\}
  \mathrm{d}t'
  \right]
  a_{n'}(t),
\end{align}
where $\bm{X}_{nn'}(\bm{k})$ is defined by
\begin{align}
  \bm{X}_{nn'}(\bm{k})
  =
  -
  i
  \int
  u^*_{n\bm{k}}(z)
  \frac{\partial}{\partial\bm{k}}
  u_{n'\bm{k}}(z)
  \,
  \mathrm{d}z.
\end{align}
From Eqs.~(\ref{eq:2DEG:En}) and (\ref{eq:2DEG:un}), the KI equation
is reduced to
\begin{align}
  \frac{\mathrm{d}a_n(t)}{\mathrm{d}t}
  =
  -
  i
  \sum_{n'}
  \Omega_{nn'}
  \mathrm{e}^{i\omega_{nn'}t}
  a_{n'}(t),
  \label{eq:2DEG:KI}
\end{align}
where
\begin{align}
  \Omega_{nn'}
  &
  =
  \frac{e\bm{F}\cdot\bm{\gamma}}{\hbar}
  \langle\zeta_{n}|z|\zeta_{n'}\rangle,
  \label{eq:2DEG:Omega}
\end{align}
and
\begin{align}
  \omega_{nn'}
  &
  = 
  \frac{E_n - E_{n'}}{\hbar}.
\end{align}

The ODEM-induced inter-subband transitions can occur during free flight
according to the KI equation of Eq.~(\ref{eq:2DEG:KI}) for
$w_{13} \neq 0$ or $w_{23} \neq 0$.  Consider the case where an
electron is in the state of $(n_0, \bm{k}_0)$ at $t = 0$.  The time
evolution of the probability coefficients $a_{n}(t)$ is determined
through the KI equation under the initial conditions
$a_{n}(0) = \delta_{n, n_0}$.  Since the probability of finding the
electron in a subband $n$ at time $t$ is $P_n(t) = |a_{n}(t)|^2$, an
effective dispersion relation $\bar{E}(\bm{k})$ may be defined by
\begin{align}
  \bar{E}_{n_0\bm{k}_0}(\bm{k}(t))
  =
  \sum_{n}
  |a_n(t)|^2
  E_{n\bm{k}(t)},
\end{align}
together with Eq.~(\ref{eq:2DEG:accel}).  Here, we add the subscripts,
$n_0$ and $\bm{k}_0$, to indicate that the effective dispersion
$\bar{E}(\bm{k})$ depends on the initial conditions.

In an initial short time interval, the probability coefficients
$a_{n}(t)$ may be approximated to
\begin{align}
  \bm{a}(t) \approx \bm{a}(0) -i \Omega \bm{a}(0) t,
  \label{eq:2DEG:ant_approx_vec}
\end{align}
where $\bm{a}(t) = (a_1(t), a_2(t), \ldots)^\mathrm{T}$ and $\Omega$
is a matrix whose $nn'$-element is given by $\Omega_{nn'}$.  For
$a_n(0) = \delta_{n, n_0}$, we have
\begin{align}
  a_n(t) = -i \Omega_{n n_0} t, \quad (n \neq n_0),
  \label{eq:2DEG:ant_approx}
\end{align}
and $|a_{n_0}(t)|^2 = 1 - \sum_{n \neq n_0} |a_{n}(t)|^2$.  The
effective dispersion is then written as
\begin{align}
  \bar{E}_{n_0 \bm{k}_0}(\bm{k}(t))
  &
  \approx
  \sum_{n \neq n_0}
  |\Omega_{n n_0}t|^2
  [E_{n\bm{k}(t)} - E_{n_0\bm{k}(t)}]
  +
  E_{n_0\bm{k}(t)}
  \nonumber
  \\
  &
  =
  [
  \delta\bm{k}(t)
  \cdot
  \bm{\gamma}
  ]^2
  \sum_{n \neq n_0}
  |
  \langle\zeta_{n}|z|\zeta_{n_0}\rangle
  |^2
  [E_{n} - E_{n_0}]
  \nonumber
  \\[-1ex]
  &
  \quad
  +
  E_{n_0\bm{k}(t)},
  \label{eq:2DEG:en0k0}
\end{align}
where $\delta \bm{k}(t) = \bm{k}(t) - \bm{k}_0 = -e\bm{F} t / \hbar$.
Using the sum rule,\cite{Wang1999} we have the following relation:
\begin{align}
  &
  \sum_{n}
  |
  \langle\zeta_{n}|z|\zeta_{n_0}\rangle
  |^2
  [E_{n} - E_{n_0}]
  \nonumber
  \\
  &
  \qquad
  =
  \mbox{$\frac{1}{2}$}
  \langle\zeta_{n_0}|[z, [H_z, z]]|\zeta_{n_0}\rangle
  =
  \mbox{$\frac{1}{2}$}
  \hbar^2 w_{33}.
\end{align}
Substituting this into Eq.~(\ref{eq:2DEG:en0k0}), we obtain the
effective dispersion for an initial short time interval:
\begin{align}
  \bar{E}_{n_0 \bm{k}_0}(\bm{k}(t))
  =
  E_{\mathrm{3D}}(\delta \bm{k}(t))
  +
  \delta \bm{k}(t)
  \cdot
  \frac{\partial E_{n_0 \bm{k}}}{\partial \bm{k}}
  \bigg|_{\bm{k} = \bm{k}_0}
  \!\!\!\!\!\!
  +
  E_{n_0 \bm{k}_0}.
\end{align}
Since $E_{\mathrm{3D}}(\bm{k})$ is the bulk dispersion and $E_{n \bm{k}}$
is the 2DEG dispersion given by Eqs.~(\ref{eq:2DEG:En}) and
(\ref{eq:2DEG:Ek}), this equation indicates that
$\bar{E}_{n_0 \bm{k}_0}(\bm{k}(t))$ is identical to the bulk
dispersion with a shifted origin for an initial short time interval,
i.e.\ the initial acceleration of a 2D electron effectively follows
the bulk dispersion relation.  The time range over which it follows
will be discussed in Sec.~\ref{sec:results}, where numerical
calculations are presented.

\subsection{Acceleration of 1DEG}

The theory for 1DEG can be derived in a parallel way to that for 2DEG,
and is briefly described in this subsection.

First, we consider 1D electronic states in a rectangle nanowire (or
nanosheet) in the absence of an applied electric field, in which
electrons move freely along the $x$-direction and are confined in the
$(y, z)$-plane by a potential $V(y, z)$.  We have the following
effective-mass equation:
\begin{align}
  \biggl[
  \frac{1}{2}
  \sum_{ij}
  w_{ij} p_i p_j
  +
  V(y, z)
  \biggr]
  \phi(x, y, z)
  =
  E
  \phi(x, y, z),
  \label{eq:1DEG:waveeq}
\end{align}
together with the boundary conditions of $\phi(x, y, z) = 0$ at the
nanowire surface.  Bescond \textit{et al.\@} \cite{Bescond2007} have
shown that the solutions of this equation can be written as follows:
\begin{align}
  E_{n k_x}
  &
  =
  E_n
  +
  E_{\mathrm{1D}}(k_x),
  \label{eq:1DEG:Enk}
  \\
  \phi_{n k_x}(x, y, z)
  &
  =
  \frac{1}{\sqrt{L}} \mathrm{e}^{ik_xx} u_{nk_x}(y, z)
  \label{eq:1DEG:phink}
\end{align}
with
\begin{align}
  u_{nk_x}(y, z)
  &
  =
  \mathrm{e}^{-ik_x(\alpha y + \beta z)}
  \zeta_n(y, z),
  \label{eq:1DEG:unk}
\end{align}
where $n$ ($=1, 2, 3, \ldots$) is the subband index associated with
the quantized motion in the $(y, z)$-plane, $k_x$ is a wavenumber
along the $x$-direction, $L$ is the length of the system,
\begin{align}
  \alpha
  &
  =
  \frac{w_{12} w_{33} - w_{23} w_{13}}{w_{22} w_{33} - w_{23}^2},
\end{align}
and
\begin{align}
  \beta
  &
  =
  \frac{w_{13} w_{22} - w_{23} w_{12}}{w_{22} w_{33} - w_{23}^2}.
\end{align}
In Eqs.~(\ref{eq:1DEG:Enk}) and (\ref{eq:1DEG:unk}), the subband level
$E_n$ and the 2D wavefunction $\zeta_n(y, z)$ are given by the
solutions of the following 2D effective-mass equation:
\begin{align}
  &
  \left\{
  \mbox{$\frac{1}{2}$}
  [
  w_{22}
  p_y^2
  +
  2
  w_{23}
  p_y
  p_z
  +
  w_{33}
  p_z^2
  ]
  +
  V(y, z)
  \right\}
  \zeta_n(y, z)
  \nonumber
  \\
  &
  \qquad
  =
  E_n
  \zeta_n(y, z).
\end{align}
The dispersion of 1DEG, $E_{\mathrm{1D}}(k_x)$, along the transport
direction is given by
\begin{align}
  E_{\mathrm{1D}}(k_x)
  =
  E_{\mathrm{3D}}(k_x)
  -
  \mbox{$\frac{1}{2}$}
  \hbar^2
  (w_{12} \alpha + w_{13} \beta) k_x^2,
\end{align}
where $E_{\mathrm{3D}}(k_x)$ ($= \frac{1}{2} \hbar^2 w_{11} k_x^2$) is the
dispersion in the bulk along the $x$-direction.  Note that
$w_{12} \alpha + w_{13} \beta > 0$ if $w_{12} \neq 0$ or
$w_{13} \neq 0$ for positive definite $w_{ij}$ as in the Si conduction
band.

In the presence of a homogeneously applied electric field, $F_x$,
along the $x$-direction, the ODEM-induced inter-subband transitions can
occur, according to the KI equation:
\begin{align}
  \frac{\mathrm{d}a_n(t)}{\mathrm{d}t}
  =
  -
  i
  \sum_{n'}
  \Omega_{nn'}
  \mathrm{e}^{i\omega_{nn'}t}
  a_{n'}(t)
  \label{eq:1DEG:KI}
\end{align}
with
\begin{align}
  \Omega_{nn'}
  &
  =
  \frac{eF_x}{\hbar}
  \langle\zeta_{n}|\alpha y + \beta z|\zeta_{n'}\rangle.
\end{align}
Note that the KI equation for 1DEG, Eq.~(\ref{eq:1DEG:KI}), is mathematically equivalent to that for 2DEG, Eq.~(\ref{eq:2DEG:KI}). The difference in dimensionality appears only through the parameter $\Omega_{nn'}$.  In a semiclassical picture, the electron motion is confined to the plane defined by the direction of the electric field and the orientation of the electron orbitals bent by the ODEM. This may explain why the KI equations take the same form for both 2DEG and 1DEG systems.

For an electron being in the initial state $(n_0, k_0)$ at $t = 0$,
its effective dispersion is written as
\begin{align}
  \bar{E}_{n_0 k_0}(k_x(t))
  =
  \sum_n |a_n(t)|^2 E_{nk_x(t)}.
\end{align}
Note that we apply periodic boundary conditions along the transport
direction and we have the time-dependent wavenumber of
$k_x(t) = k_x(0) - eFt /\hbar$.\cite{Krieger1986}  In an initial
short period of time of the acceleration, the effective dispersion is
approximated to
\begin{align}
  \bar{E}_{n_0 k_0}(k_x(t))
  =
  E_{\mathrm{3D}}(\delta k_x(t))
  +
  \delta k_x(t) \frac{\partial E_{n_0 k_x}}{\partial k_x}\bigg|_{k_x = k_0}
  \!\!\!\!\!\!\!\!
  +
  E_{n_0 k_0},
  \label{eq:1DEG:en0k0}
\end{align}
where $\delta k_x(t) = k_x(t) - k_0 = -eF_x t / \hbar$ and we have
used the following relation:
\begin{align}
  \sum_{n}
  |
  \langle \zeta_{n} | \alpha y + \beta z | \zeta_{n_0} \rangle
  |^2
  [E_{n} - E_{n_0}]
  =
  \mbox{$\frac{1}{2}$} \hbar^2 (\alpha w_{12} + \beta w_{13}).
\end{align}

\section{Numerical Results and Discussion\label{sec:results}}

\subsection{2DEG States}

\begin{figure}[b]%[htbp]
  \centering
  \includegraphics[scale=0.35]{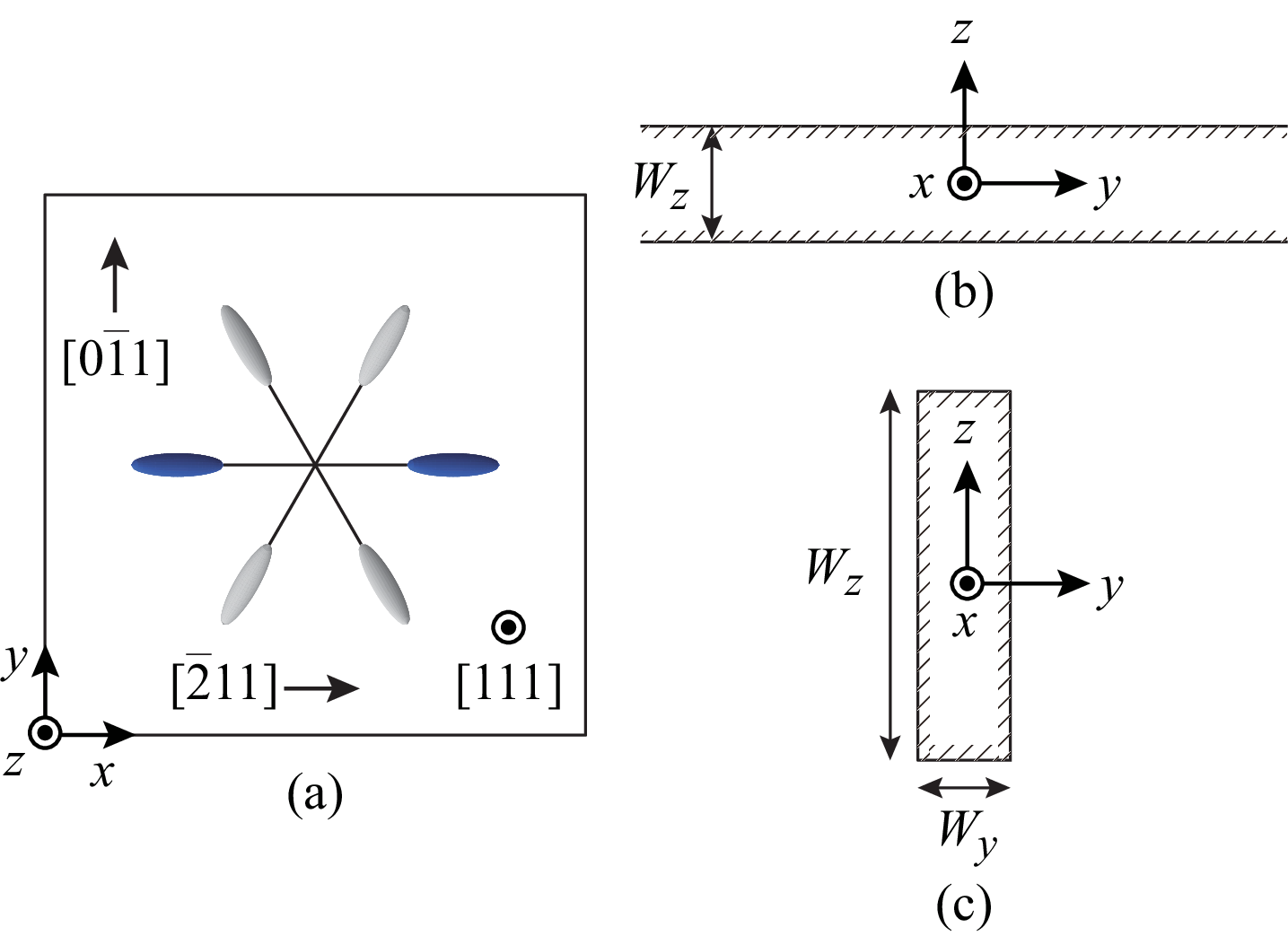}
  \caption{(a) Coordinate system for the Si(111) substrate together
    with a schematic representation of the constant-energy ellipses of
    the Si conduction band.  Numerical examples are given only for the
    valleys highlighted in blue.  (b) Slab of thickness $W_z$ relevant
    for the discussion of 2DEG motion.  (c) Fin-shape structure of
    size $W_y \times W_z$ relevant for the discussion of 1DEG
    motion.  \label{fig:system111}}
\end{figure}

As a first example, we consider 2DEG in a Si slab of thickness $W_z$
with the (111) surface, whose schematic diagram is shown in
Figs.~\ref{fig:system111}(a) and (b).  As shown in
Fig.~\ref{fig:system111}(a), the $x$, $y$, and $z$ axes are defined
along the [\={2}11], [0\={1}1], and [111] directions, respectively.
For the sake of simplicity, the confinement potential is assumed to be
an infinite quantum-well type of
\begin{align}
  V(z)
  =
  \begin{cases}
    0 & (|z| < \frac{1}{2} W_z)
    \\
    \infty & (\mbox{otherwise})
  \end{cases}.
  \label{eq:2DEG:Vwell}
\end{align}
We consider electrons in the [100] valleys (highlighted in blue in
Fig.~\ref{fig:system111}(a)), whose reciprocal effective-mass tensor
is given by
\begin{align}
  w
  =
  \begin{bmatrix}
    \displaystyle
    \frac{1}{3 m_t} + \frac{2}{3 m_l} 
    &
      \displaystyle
      0
    & 
      \displaystyle
      \frac{\sqrt{2}}{3 m_t} - \frac{\sqrt{2}}{3 m_l}
    \\
    \displaystyle
    0
    & 
      \displaystyle
      \frac{1}{m_t}
    & 
      \displaystyle
      0
    \\
    \displaystyle
    \frac{\sqrt{2}}{3m_t} - \frac{\sqrt{2}}{3m_l}
    & 
      \displaystyle
      0
    & 
      \displaystyle
      \frac{2}{3m_t} + \frac{1}{3m_l}
  \end{bmatrix},
  \label{eq:w111}
\end{align}
where $m_l = 0.916 \, m_0$ and $m_t = 0.19 \, m_0$ are the
longitudinal and transverse effective-masses in the bulk Si,
respectively.  Since $w_{13} \neq 0$, the ODEM-induced inter-subband
transitions can occur for $|F_x| > 0$.

\begin{figure}[htbp]
  \centering
  \includegraphics[scale=0.475]{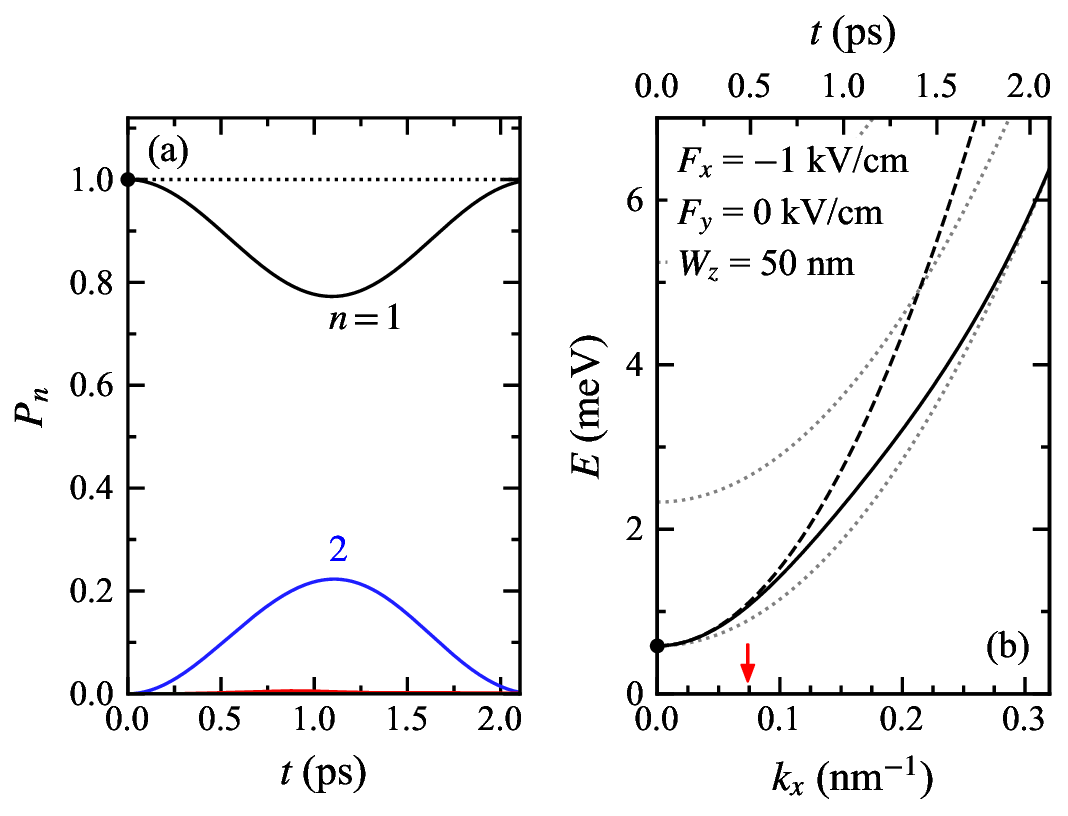}
  \caption{(a) Time evolution of the probability coefficients $P_n(t)$
    in a Si slab of thickness $W_z = 50 \, \mathrm{nm}$ with the (111)
    surface for an applied electric field $F_x$ ($\parallel$[\={2}11])
    $ = -1 \, \mathrm{kV/cm}$ and $F_y$ ($\parallel$[0\={1}1]) $= 0$.
    The solid marker represents the initial condition of
    $a_n(0) = \delta_{n, 1}$. (b) Effective dispersion
    $\bar{E}(\bm{k}(t))$ with the initial conditions of
    $(n, \bm{k}) = (1, \bm{0})$ (solid line) together with the bulk
    dispersion offset by the subband level $E_1$,
    $E_{\mathrm{3D}}(\bm{k}(t)) + E_1$, (dashed line) and the 2DEG
    dispersions, $E_n + E_{\mathrm{2D}}(\bm{k}(t))$ (dotted lines).
    The red arrow shows the critical wavenumber $k_\mathrm{c}$
    ($= -e F_x t_\mathrm{c} / \hbar$) calculated from
    Eq.~(\ref{eq:2DEG:tc}). \label{fig:2DEG:ref}}
\end{figure}

We numerically solve the KI equation of Eq.~(\ref{eq:2DEG:KI}) to
obtain the time dependence of $P_n(t) = |a_n(t)|^2$ with the initial
conditions of $a_{n}(0) = \delta_{n, 1}$ and $\bm{k}(0) = \bm{0}$.
The results are shown in Fig.~\ref{fig:2DEG:ref}(a) for
$W_z = 50 \, \mathrm{nm}$, $F_x = -1 \, \mathrm{kV/cm}$, and
$F_y = 0$.  Focusing on $P_2(t)$, we can see that it gradually
increases as time passes from $t = 0$, reaches a maximum value at
$t \approx 1 \, \mathrm{ps}$, then decreases, and reaches 0 again at
$t \approx 2 \, \mathrm{ps}$.  The gradual increase at around $t = 0$
corresponds to the ODEM-induced inter-subband transition.  
The oscillations between subbands result from phase interference effects and are fundamentally equivalent to coherent transitions in a two-level quantum system.
If we
consider only the lowest two subbands, $n=1$ and $n=2$, we can solve
the KI equation analytically and obtain the following equation:
\begin{align}
  P_2(t)
  =
  \frac{2\Omega_{21}^2}{f^2}
  (1 - \cos ft),
  \label{eq:2DEG:P2}
\end{align}
where
$f = [(\Omega_{22} - \Omega_{11} + \omega_{21})^2 + 4
\Omega_{21}^2]^{1/2}$.  For the infinite quantum-well model of
Eq.~(\ref{eq:2DEG:Vwell}), $\Omega_{22} = \Omega_{11}$ and $f$ is
reduced to $f = (\omega_{21}^2 + 4 \Omega_{21}^2)^{1/2}$.
We see that the maximum value of Eq.~(\ref{eq:2DEG:P2}) is determined
by the ratio of the field-induced term $\Omega_{21}$ to the subband
spacing term $\omega_{21}$. On the other hand, in the limit of
$t = 0$, Eq.~(\ref{eq:2DEG:P2}) becomes $P_2(t) = (\Omega_{12} t)^2$,
which is consistent with Eq.~(\ref{eq:2DEG:ant_approx}), and $P_2(t)$
is determined only by $\Omega_{12}$.

We calculate the effective dispersion,
$\bar{E}_{1,\bm{0}}(\bm{k}(t))$, from $P_n(t)$ shown in
Fig.~\ref{fig:2DEG:ref}(a) for $W_z = 50 \, \mathrm{nm}$,
$F_x = -1 \, \mathrm{kV/cm}$, and $F_y = 0$.  The result is plotted in
Fig.~\ref{fig:2DEG:ref}(b) by the solid line, in which we also plot
the bulk dispersion offset by the subband level $E_1$,
$E_{\mathrm{3D}}(\bm{k}(t)) + E_1$, by the dashed line and the 2DEG
dispersions, $E_n + E_{\mathrm{2D}}(\bm{k}(t))$, by the dotted lines
for comparison.  We see that the effective dispersion follows the bulk
dispersion for an initial interval
$0 < t < t_\mathrm{c} \approx 0.5 \, \mathrm{ps}$, which corresponds
to
$0 < k_x < k_\mathrm{c} \; (= -e F_x t_\mathrm{c} / \hbar) \approx
0.07 \, \mathrm{nm}^{-1}$. The critical time $t_\mathrm{c}$ in the
infinite quantum-well model may be estimated as follows.  For the
approximation of Eq.~(\ref{eq:2DEG:ant_approx_vec}) to hold, at least
the relation $\mathrm{e}^{i\omega_{nn'} t} \approx 1$, i.e.\
$\omega_{nn'}t \ll 1$, must hold.  Since transitions between adjacent
subbands dominate for the ODEM-induced inter-subband transitions, the
critical time $t_\mathrm{c}$ is assumed to be estimated by
$t_\mathrm{c} = \frac{1}{2}(\omega_{n_{\mathrm{c}} + 1,
  n_{\mathrm{c}}}^{-1} + \omega_{n_{\mathrm{c}},
  n_{\mathrm{c}}-1}^{-1})$, where $n_\mathrm{c}$ is the subband index
at which the effective energy is equal to the 2DEG energy at
$t = t_\mathrm{c}$, i.e.\ it is given by the solution of
$E_{\mathrm{3D}}(\bm{k}(t_\mathrm{c})) + E_1 =
E_{\mathrm{2D}}(\bm{k}(t_\mathrm{c})) + E_{n_\mathrm{c}}$.  We then
obtain the critical time in the infinite quantum-well model:
\begin{align}
  t_\mathrm{c}
  =
  [(\tau_0^4 + \tau_1^4)^{1/2} - \tau_1^2]^{1/2},
  \label{eq:2DEG:tc}
\end{align}
where $\tau_0$ and $\tau_1$ are defined as
\begin{align}
  \tau_0
  =
  \left|
  \frac{W_z}{\pi w_{33} e\bm{F}\cdot\bm{\gamma}}
  \right|^{1/2},
  \quad
  \tau_1
  =
  \left|\frac{\pi \hbar}{\sqrt{2} e\bm{F}\cdot\bm{\gamma}W_z}\right|.
\end{align}
The red arrow in Fig.~\ref{fig:2DEG:ref}(b) shows
$k_c = - eF_x t_\mathrm{c} / \hbar$.  Since
$\tau_0 / \tau_1 = |2e\bm{F}\cdot\bm{\gamma} W_z^3 / \pi^3 \hbar^2
w_{33}|^{1/2}$, $\tau_\mathrm{c} \approx \tau_0$ \cite{kc} for wider
$W_z$ or stronger $\bm{F}$.

\begin{figure}[htbp]
  \centering
  \includegraphics[scale=0.475]{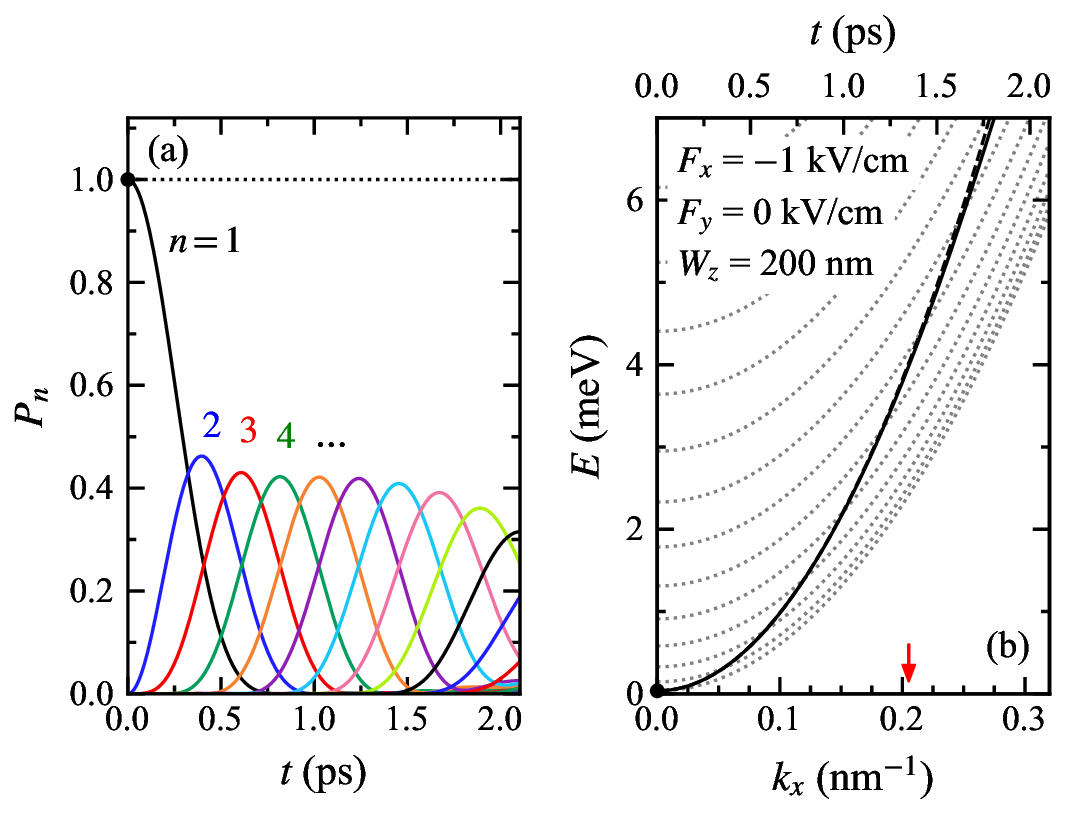}
  \caption{The same as Fig.~\ref{fig:2DEG:ref} but for a thicker slab
    of $W_z = 200 \, \mathrm{nm}$. \label{fig:2DEG:thicker}}
\end{figure}

As the slab thickness $W_z$ increases, the subband spacing decreases,
resulting in a longer initial interval.  Fig.~\ref{fig:2DEG:thicker}
shows the same results as in Fig.~\ref{fig:2DEG:ref}, but with the
thickness $W_z$ increased to $200 \, \mathrm{nm}$.  We see that,
unlike the thinner case of Fig.~\ref{fig:2DEG:ref}(a), $P_n(t)$ peaks
in the order of $n = 2, 3, 4, \ldots$ as time passes.  We also see
that the effective dispersion coincides with the bulk dispersion up to
$t_\mathrm{c} \approx 1.3 \, \mathrm{ps}$, which corresponds to
$k_\mathrm{c} \approx 0.2 \, \mathrm{nm}^{-1}$.  As $W_z$ is further
thickened, the initial period becomes very long according to
Eq.~(\ref{eq:2DEG:tc}). This implies that in the thick limit, the 2DEG
transport properties may approach those of the bulk.

\begin{figure}[htbp]
  \centering
  \includegraphics[scale=0.475]{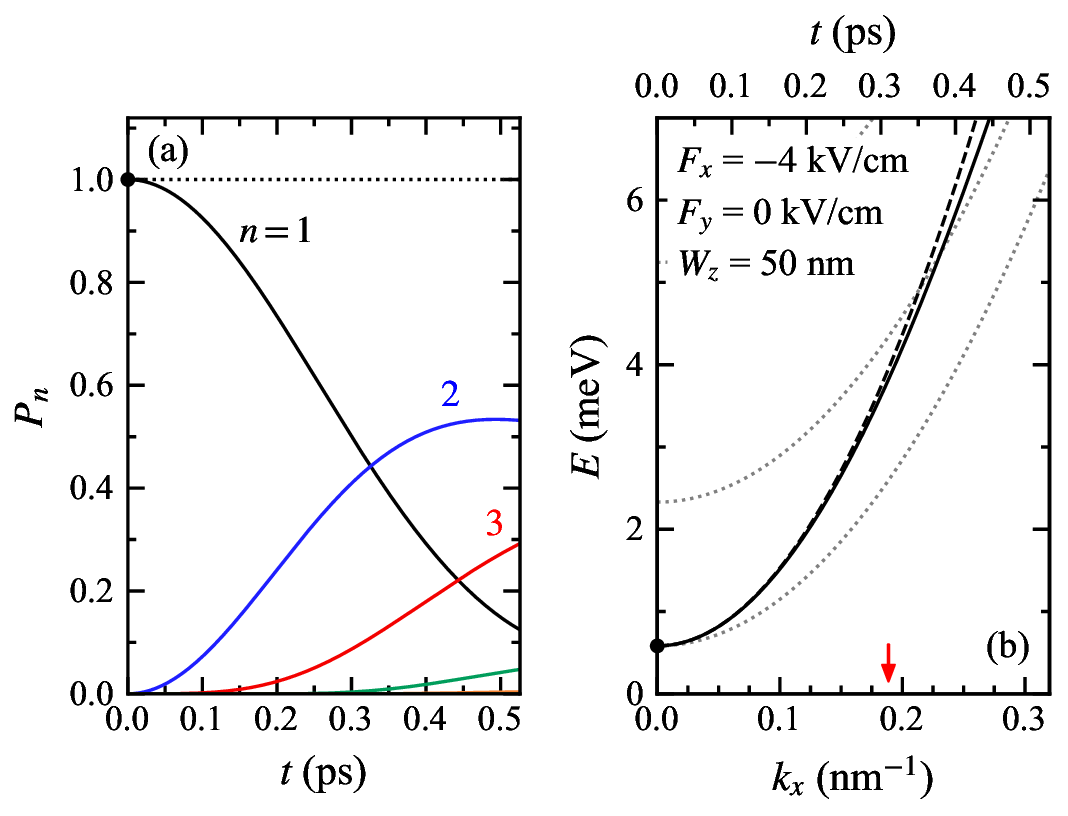}
  \caption{The same as Fig.~\ref{fig:2DEG:ref} but for a higher
    applied electric field of $|F_x| = 4 \,
    \mathrm{kV/cm}$. \label{fig:2DEG:highF}}
\end{figure}

The applied electric field enhances the ODEM-induced inter-subband
transitions because $\Omega$ is proportional to the field.
Fig.~\ref{fig:2DEG:highF} shows the same results as in
Fig.~\ref{fig:2DEG:ref}, but with the applied electric field $|F_x|$
increased to $4 \, \mathrm{kV/cm}$.  As can be seen by comparing
Fig.~\ref{fig:2DEG:highF}(b) with Fig.~\ref{fig:2DEG:ref}(b),
increasing the field strength by a factor of four increases the
critical wavenumber, $k_\mathrm{c}$, by a factor of $\approx 2.5$, and
the effective dispersion relation follows the bulk dispersion up to
higher energies.  As the electric field increases, the critical time
$t_\mathrm{c}$ decreases. However, $t_\mathrm{c}$ ($\approx \tau_0$)
decreases with $F$ to the $-1/2$ power, whereas acceleration in the
$k$-space follows $F$ to the first power, so the critical wave number
$k_\mathrm{c}$ increases with the electric field.

\subsection{1DEG States}

\begin{figure}[htbp]
  \centering
  \includegraphics[scale=0.475]{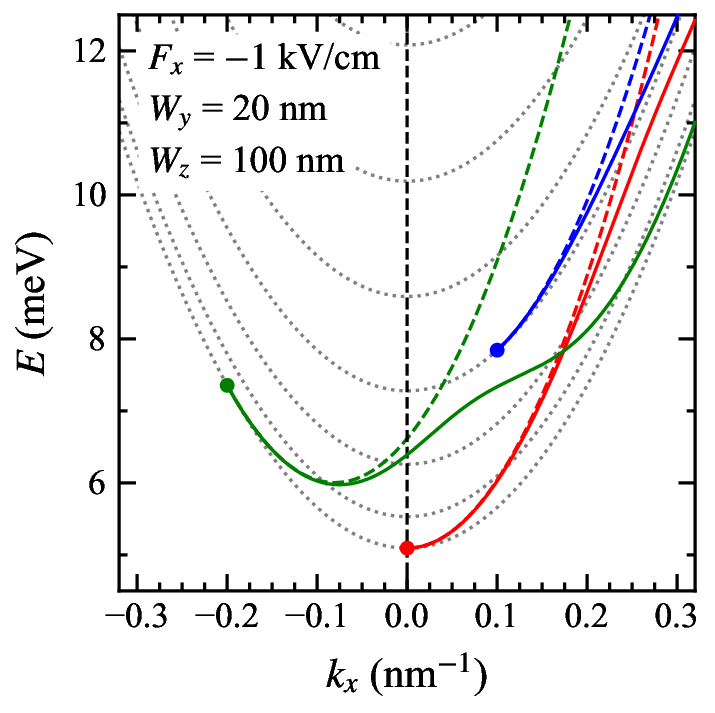}
  \caption{Solid lines show the effective dispersions
    $\bar{E}_{n_0 k_0}(k_x(t))$ of 1DEG in a
    $20 \, \mathrm{nm} \times 100 \, \mathrm{nm}$ nanowire on the
    Si(111) substrate (see Figs.~\ref{fig:system111}(a) and (c)) for
    an applied electric field $F_x = -1 \, \mathrm{kV/cm}$. The
    results of three different initial conditions, $(n_0, k_0)$, are
    plotted: $(n_0, k_0) = (1, \, 0\, \mathrm{nm}^{-1})$ (red),
    $(1, \, -0.2 \, \mathrm{nm}^{-1})$ (green), and
    $(4, \, 0.1 \, \mathrm{nm}^{-1})$ (blue).  Dotted lines show the
    1DEG dispersions and dashed lines the shifted bulk dispersions of
    Eq.~(\ref{eq:1DEG:en0k0}). \label{fig:1deg:111}}
\end{figure}

Next, we show the results for 1DEG systems.  Fig.~\ref{fig:1deg:111}
shows the effective dispersions, $\bar{E}_{n_0 k_0}(k_x)$, of 1DEG in
a rectangle nanowire on the Si(111) substrate, whose schematic
diagram is given in Figs.~\ref{fig:system111}(a) and (c), for
$F_x = -1 \, \mathrm{kV/cm}$, $W_y = 20 \, \mathrm{nm}$, and
$W_z = 100 \, \mathrm{nm}$.  The confinement potential is assume to be
\begin{align}
  V(y, z)
  =
  \begin{cases}
    0 & (|y| < \frac{1}{2} W_y \mbox{~and~}
        |z| < \frac{1}{2} W_z)
    \\
    \infty & (\mbox{otherwise})
  \end{cases}.
  \label{eq:1deg:Vconf}
\end{align}
The reciprocal effective-mass tensor is given by Eq.~(\ref{eq:w111}),
and $w_{13} \neq 0$, leading to the ODEM-induced transitions for
$|F_x| > 0$.  We plot $\bar{E}_{n_0 k_0}(k_x)$ for three different
initial conditions: $(n_0, k_0) = (1, \, 0 \, \mathrm{nm}^{-1})$
(represented by the red marker in Fig.~\ref{fig:1deg:111}),
$(1, \, -0.2 \, \mathrm{nm}^{-1})$ (green), and
$(4, \, 0.1 \, \mathrm{nm}^{-1})$ (blue).  Focusing on the initial
condition of the green marker, for example, we can clearly see that
for a short initial period, $\bar{E}_{n_0 k_0}(k_x(t))$ follows the
shifted bulk dispersion and then approaches the 1DEG dispersion.

So far, we have focused on the transition from low-dimensional to bulk
systems, showing mainly the results of calculations under weak
electric field conditions for relatively large systems.  In
state-of-the-art MOS transistors, the short channel length may result
in a relatively high electric field region.  Under such conditions,
the effect of ODEM-induced inter-subband transitions may not be
ignored, even in ultra-small devices.

\begin{figure}[htbp]
  \centering
  \includegraphics[scale=0.35]{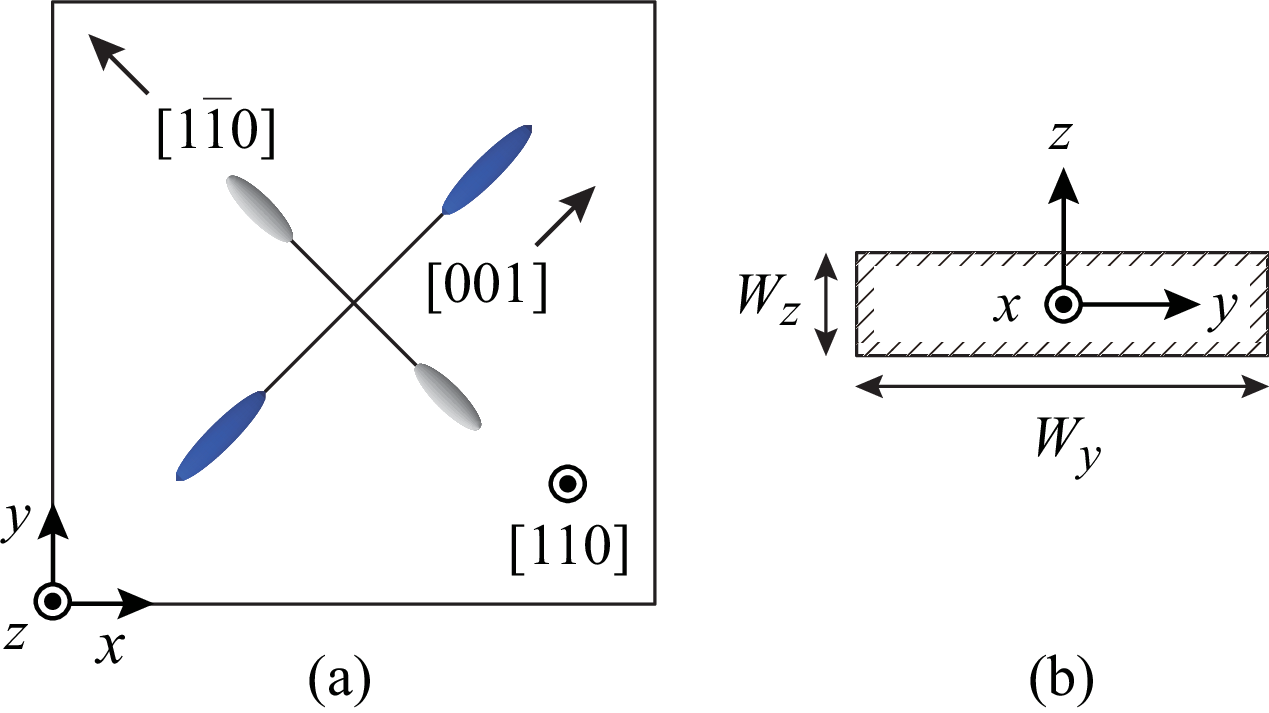}
  \caption{(a) Geometry relevant to the discussion of the Si(110)
    substrate together with schematic representation of the
    constant-energy ellipses of the Si conduction band.  Numerical
    examples are given only for the valleys highlighted in blue. (b)
    Nanosheet-shape structure with size $W_y \times
    W_z$. \label{fig:system110}}
\end{figure}

\begin{figure}[htbp]
  \centering
  \includegraphics[scale=0.475]{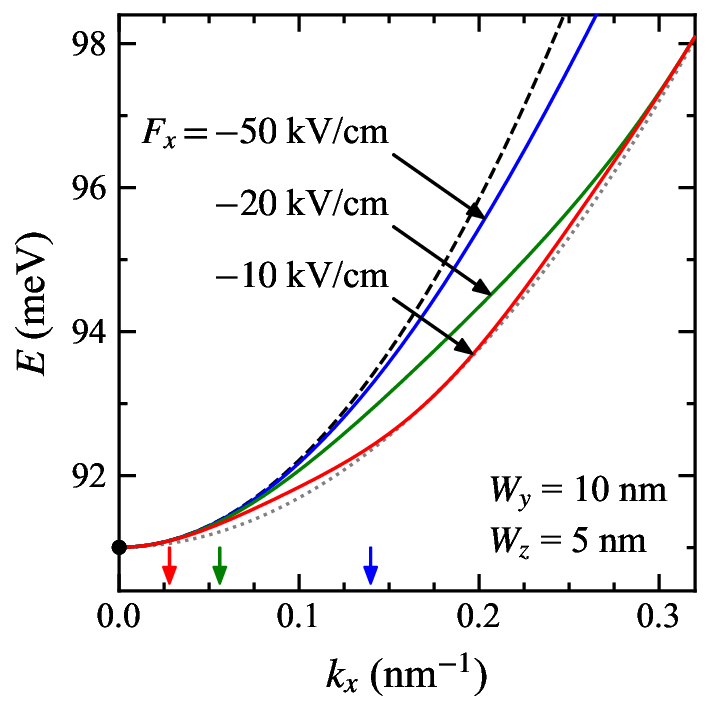}
  \caption{Solid lines show the effective dispersions
    $\bar{E}_{n_0 k_0}(k_x(t))$ of 1DEG in a
    $10 \, \mathrm{nm} \times 5 \, \mathrm{nm}$ nanowire on the
    Si(110) substrate (see Fig.~\ref{fig:system110}) for three
    different applied electric fields, $F_x = -10 \, \mathrm{kV/cm}$
    (red line), $-20 \, \mathrm{kV/cm}$ (green), and
    $-50 \, \mathrm{kV/cm}$ (blue).  Dotted line shows the 1DEG
    dispersion and dashed line the shifted bulk dispersions of
    Eq.~(\ref{eq:1DEG:en0k0}). Small vertical arrows indicate the
    critical wavenumbers of
    $k_\mathrm{c} = -eF_x / \hbar \omega_{12}$. \label{fig:1deg:110}}
\end{figure}

We consider a rectangle nanowire on the Si(110) substrate, whose
schematic diagram is given in Fig.~\ref{fig:system110}.  As shown in
Fig.~\ref{fig:system110}(a), the $x$-axis is chosen to be at a
45-degree angle to the [001] direction on the (110) surface and
electrons in the [001] valleys (highlighted in blue in the figure) are
considered.  The reciprocal effective-mass tensor is then given by
\begin{align}
  w
  =
  \begin{bmatrix}
    \displaystyle
    \phantom{-}
    \frac{1}{2 m_t} + \frac{1}{2 m_l}
    &
      \displaystyle
      -\frac{1}{2 m_t} + \frac{1}{2 m_l}
    & 
      \displaystyle
      0
    \\[2ex]
    \displaystyle
    -\frac{1}{2 m_t} + \frac{1}{2 m_l}
    & 
      \displaystyle
    \phantom{-}
      \frac{1}{2 m_t} + \frac{1}{2 m_l}
    & 
      \displaystyle
      0
    \\[2ex]
    \displaystyle
    0 
    & 
      \displaystyle
      0 
    & 
      \displaystyle
      \frac{1}{m_t}
  \end{bmatrix},
  \label{eq:w110}
\end{align}%
and $w_{12} \neq 0$, leading to the ODEM-induced transitions for
$|F_x| > 0$.  Fig.~\ref{fig:1deg:110} shows the effective dispersions,
$\bar{E}_{n_0 k_0}(k_x)$, of 1DEG in the rectangle nanowire with
$W_y = 10 \, \mathrm{nm}$ and $W_z = 5 \, \mathrm{nm}$ for three
different applied electric fields, $F = -10 \, \mathrm{kV/cm}$ (red
line), $-20 \, \mathrm{kV/cm}$ (green), and $-50 \, \mathrm{kV/cm}$
(blue).  The confinement potential is assumed to be given by
Eq.~(\ref{eq:1deg:Vconf}).  It can be seen that even for a very small
structure, the effect of ODEM-induced inter-subband transitions cannot
be neglected under the relatively high electric field conditions.  Due
to the smaller cross section of the structure, the confinement energy
is so large that the two-level model discussed around
Eq.~(\ref{eq:2DEG:P2}) could be a good approximation and the critical
wavenumber $k_\mathrm{c}$ may be written as
$k_\mathrm{c} \approx -eF_x / \hbar f \approx -eF_x / \hbar
\omega_{12}$.  Here, $\hbar \omega_{21} = E_2 - E_1$ is the subband
spacing and we have assumed that $\omega_{21} \gg |\Omega_{21}|$.  The
small vertical arrows in Fig.~\ref{fig:1deg:110} indicate
$k_\mathrm{c} = -eF_x / \hbar \omega_{12}$.  Up to $k_\mathrm{c}$
obtained in this way, the effective dispersion relation is in close
agreement with the shifted bulk dispersion relation.

\subsection{2DEG Drift Velocity}

To complement the preceding discussion of energy and subband dynamics, it is highly desirable to evaluate how bulk-like behavior manifests in transport-relevant observables such as the drift velocity. Here, we present the drift velocity, $\langle \bm{v} \rangle$, of 2DEG, calculated using a simplified path integration method.\cite{Chambers1952,Esaki1970}

For the 2DEG state
$\psi(\bm{x}, z, t)$ of Eq.~(\ref{eq:2DEG:psi}), the expectation value
of the electron velocity, $\bar{\bm{v}}(t)$, is given by
\cite{Iafrate2020}
\begin{align}
  \bar{\bm{v}}(t)
  =
  &
  \int
  \psi^*(\bm{x}, z, t)
  \hat{\bm{v}}
  \psi(\bm{x}, z, t)
  \mathrm{d}{\bm{x}}
  \mathrm{d}{z}
  % \nonumber
  % \\
  % =
  % &
  % \sum_{n'n}
  % a^*_{n'}(t)
  % a_{n}(t)
  % \mathrm{e}^{-i\omega_{nn'}t}
  % \nonumber
  % \\[-1ex]    
  % &
  % \times
  % \int
  % {\phi}'^*_{n'\bm{k}(t)}(\bm{x}, z, t)
  % \hat{\bm{v}}
  % % \frac{i}{\hbar}[H' , \bm{x}]
  % \phi'_{n\bm{k}(t)}(\bm{x}, z, t)
  % \mathrm{d}{\bm{x}}
  % \mathrm{d}{z}
  \nonumber
  \\
  =
  &
  \sum_{n}
  |a_n(t)|^2 \bm{v}_n(\bm{k}(t)) + V_2(t) \bm{\gamma},
\end{align}
where $\hat{\bm{v}} = (i/\hbar)[H' , \bm{x}]$,
$\bm{v}_n(\bm{k}) = \hbar^{-1} \partial E_{n\bm{k}} / \partial
\bm{k}$, and
\begin{align}
  V_2(t)
  &
  =
  i
  \sum_{n'n}
  a^*_{n'}(t)
  a_{n}(t)
  \mathrm{e}^{i\omega_{n'n}t}
  \omega_{n'n}
  \langle\zeta_{n'}|z|\zeta_{n}\rangle.
\end{align}
Since
$E_{n\bm{k}} = E_{n} + E_{\mathrm{2D}}(\bm{k}) = E_{n} +
E_{\mathrm{3D}}(\bm{k}) - \frac{1}{2} \hbar^2 w_{33} (\bm{\gamma}
\cdot \bm{k})^2$, we have
\begin{align}
  \bar{\bm{v}}(t)
  &
  =
  \bm{v}_{\mathrm{2D}}(t)
  +
  V_2(t) \bm{\gamma} 
  \nonumber
  \\
  &
  =
  \bm{v}_{\mathrm{3D}}(t)
  +
  [V_1(t) + V_2(t)] \bm{\gamma},
\end{align}
where $V_1(t) = - \hbar w_{33} \bm{\gamma}\cdot\bm{k}(t)$, and
$\bm{v}_{\mathrm{3D}}$ and $\bm{v}_{\mathrm{2D}}$ are the group
velocities given by the dispersion relations $E_\mathrm{3D}(\bm{k})$
and $E_\mathrm{2D}(\bm{k})$, respectively:
\begin{align}
  \bm{v}_{\mathrm{3D}}(t)
  &
  =
  \frac{1}{\hbar}
  \frac{\partial E_{\mathrm{3D}}(\bm{k})}{\partial \bm{k}}
  =
  \hbar
  \begin{bmatrix}
    w_{11} & w_{12} \\
    w_{12} & w_{22}
  \end{bmatrix}
  \bm{k}(t),
  \\
  \bm{v}_{\mathrm{2D}}(t)
  &
  =
  \frac{1}{\hbar}
  \frac{\partial E_{\mathrm{2D}}(\bm{k})}{\partial \bm{k}}
  =
  \bm{v}_{\mathrm{3D}}(t)
  +
  V_1(t)\bm{\gamma}.
\end{align}
The drift velocity $\langle \bm{v} \rangle$, taking into account the
scattering time $\tau$, is then written as
\begin{align}
  \langle \bar{\bm{v}}\rangle
  =
  \frac{1}{\tau}
  \int_0^{\infty}
  \bar{\bm{v}}(t)
  \mathrm{e}^{-t/\tau}
  \mathrm{d}t.
\end{align}
According to the velocity expectation value $\bar{\bm{v}}(t)$, which
can be decomposed as
$\bar{\bm{v}}(t) = \bm{v}_{\mathrm{2D}}(t) +  V_2(t) \bm{\gamma}
= \bm{v}_{\mathrm{3D}}(t) + [V_1(t) + V_2(t)] \bm{\gamma}$,
we decompose the drift velocity $\langle \bar{\bm{v}}\rangle$ as
follows:
\begin{align}
  \langle \bar{\bm{v}}\rangle
  =
  \bm{v}_\mathrm{2D}
  +
  \bm{v}_2
  =
  \bm{v}_\mathrm{3D}
  +
  \bm{v}_1
  +
  \bm{v}_2,
\end{align}
where
$\bm{v}_\mathrm{3D} = -e \tau (w_{11} F_1 + w_{12} F_2, \; w_{22} F_2
+ w_{12} F_1)$, $\bm{v}_\mathrm{2D} = \bm{v}_\mathrm{3D} + \bm{v}_1$,
$\bm{v}_1 = e \tau w_{33} (\bm{\gamma} \cdot \bm{F})\, \bm{\gamma}$,
and
\begin{align}
  \bm{v}_2
  &
  =
  \frac{1}{\tau}
  \int_0^{\infty}
  V_2(t) \mathrm{d}t\,
  \bm{\gamma}.
\end{align}

\begin{figure}[htbp]
  \centering
  \includegraphics[scale=0.475]{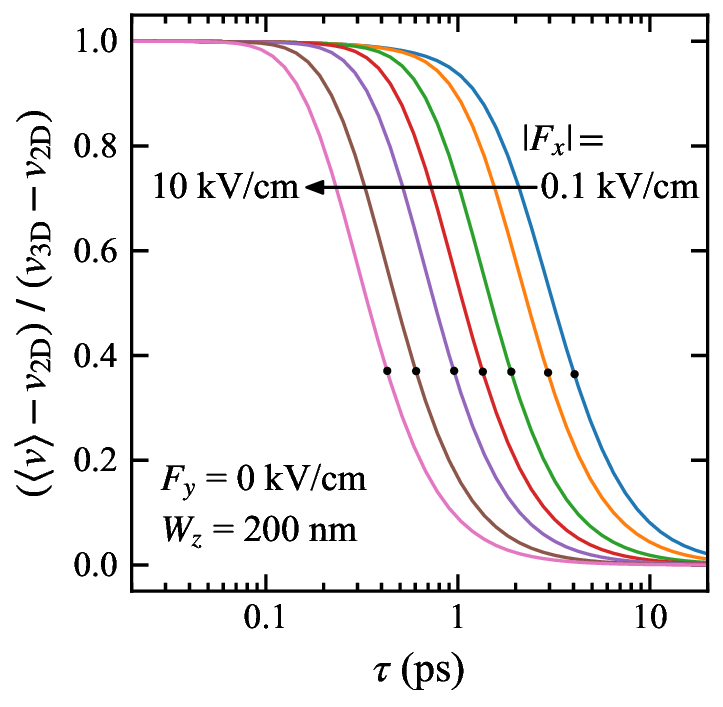}
  \caption{Scattering time $\tau$ dependence of the normalized
    $x$-component of drift velocity,
    $v_{x\mathrm{norm}} = (\langle v_x \rangle - v_{\mathrm{2D}x}) /
    (v_{\mathrm{3D}x} - v_{\mathrm{2D}x})$ in a Si slab of thickness
    $W_z = 200\,\mathrm{nm}$ with the (111) surface for applied
    electric fields $|F_x|$ ($\parallel$[\={2}11]) $= 0.1$, $0.2$,
    $0.5$, $1.0$, $2.0$, $5.0$, and $10.0\, \mathrm{kV/cm}$ and $F_y$
    ($\parallel$[0\={1}1]) $ = 0$. Black dots indicate the critical
    time, $t_\mathrm{c}$, of
    Eq.~(\ref{eq:2DEG:tc}). \label{fig:2deg:vd-tau}}
\end{figure}

We calculate the drift velocity $\langle \bar{\bm{v}} \rangle$ of the
Si slab shown in Fig.~\ref{fig:system111}(a) under applied electric
fields along the $x$-direction.  Fig.~\ref{fig:2deg:vd-tau} shows the
normalized $x$-component of drift velocity,
$v_{x\mathrm{norm}} = (\langle v_x \rangle - v_{\mathrm{2D}x}) /
(v_{\mathrm{3D}x} - v_{\mathrm{2D}x})$, as a function of the
scattering time $\tau$ for a fixed slab thickness of
$W_z = 200 \, \mathrm{nm}$.  The initial $\bm{k}(t)$ is assumed to be
$\bm{k}(0) = \bm{0}$.  We see that if the scattering time is short
enough, $v_{x\mathrm{norm}}$ approaches unity, which corresponds to
$\langle v_x \rangle = v_\mathrm{3D}$.  On the other hand, if the
scattering time is long enough, $v_{x\mathrm{norm}}$ approaches zero,
which corresponds to $\langle v_x \rangle = v_\mathrm{2D}$.  The
critical time, $t_\mathrm{c}$, of Eq.~(\ref{eq:2DEG:tc}) is marked by
black dots in the figure, indicating that the transition from
$v_\mathrm{3D}$ to $v_\mathrm{2D}$ occurs at $t \approx t_\mathrm{c}$.
We can also see that the transition time becomes shorter when the
magnitude of the applied electric field, $|F_x|$, is larger.

\begin{figure}[htbp]
  \centering
  \includegraphics[scale=0.475]{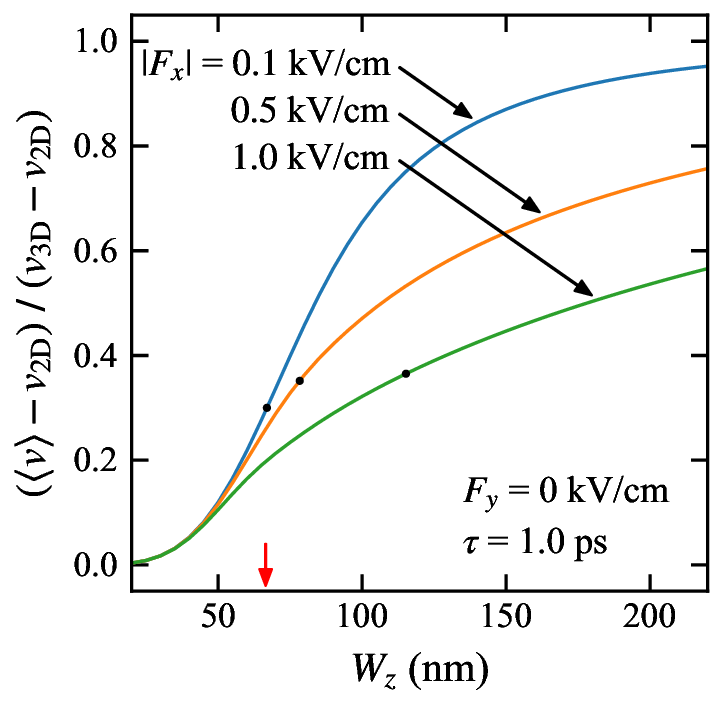}
  \caption{Slab thickness $W_z$ dependence of the normalized
    $x$-component of drift velocity $v_{x\mathrm{norm}}$ in an (111)
    surface Si slab for the scattering time
    $\tau = 1.0 \, \mathrm{ps}$.  The applied electric fields are
    $|F_x|$ ($\parallel$[\={2}11]) $= 0.1$, $0.5$, and
    $10.0\, \mathrm{kV/cm}$ and $F_y$ ($\parallel$[0\={1}1]) $ =
    0$. Black dots indicate the critical $W_z$, which is a solution of
    the equation $t_\mathrm{c}(W_z) = \tau$. 
    The red arrow shows the critical well width $W_\mathrm{c}$.
    \label{fig:2deg:vd-W}}
\end{figure}

Fig.~\ref{fig:2deg:vd-W} shows $v_{x\mathrm{norm}}$ as a function of
the slab thickness $W_z$ for a fixed scattering time
$\tau = 1.0 \, \mathrm{ps}$.  We see that $v_{x\mathrm{norm}} \to 0$
or $\langle v_x \rangle \to v_{\mathrm{2D}x}$ for thinner $W_z$.  On
the other hand, $\langle v_x \rangle$ approaches $v_{\mathrm{3D}x}$
when the slab thickness becomes wider.  These results may indicate
that the transport properties of a 2DEG system correctly approach
those of the bulk in the wider $W_z$ limit.

Now let us consider the limit in which the ODEM-induced effect vanishes. From Eq.~(\ref{eq:2DEG:Omega}), it is evident that this effect is governed by the magnitude of $\eta = \bm{F} \cdot \bm{\gamma}$. This limit can be reached by setting either the electric field ($\bm{F}$) or the ODEM ($\bm{\gamma}$) to zero. When $\eta \to 0$, the critical time $t_\mathrm{c}$, given by Eq.~(44), becomes a constant $t_\mathrm{c}|_{\eta \to 0} = W_z^2 / (\pi^2 \hbar w_{33})$, which is independent of $\eta$. Interestingly, a subtle phenomenon arises in this limit. If the scattering time $\tau$ is much shorter than $t_\mathrm{c}|_{\eta \to 0}$, the electron dynamics resemble those of the bulk. Conversely, if the scattering time $\tau$ is longer than $t_\mathrm{c}|_{\eta \to 0}$, bulk-like behavior does not emerge. Since $t_\mathrm{c}|_{\eta \to 0}$ depends on the quantum well width $W_z$, let us define $W_\mathrm{c}$ as the well width at which  $t_\mathrm{c}|_{\eta \to 0} = \tau$. This gives $W_\mathrm{c} = \pi (\hbar \tau w_{33})^{1/2}$. The location of this critical well width $W_\mathrm{c}$ is indicated by the red arrow in Fig.~\ref{fig:2deg:vd-W}. When the well width exceeds $W_\mathrm{c}$, the electron dynamics transition from two-dimensional-like to bulk-like as the electric field decreases. In this regime, ODEM-induced inter-subband transitions could potentially be observed experimentally by examining the electric field dependence of the electron mobility. However, the results presented in Fig.~\ref{fig:2deg:vd-W} are based on a simplified path integration method at absolute zero temperature within a single-valley and constant scattering-time approximation. Therefore, a more quantitative approach is necessary for direct comparison with experimental data.

\section{Conclusions\label{sec:conclusion}}

We have studied the motion of electrons under homogeneously applied
electric fields in 1DEG or 2DEG systems with non-zero off-diagonal
effective mass (ODEM).  The equation describing the time evolution of
a probability coefficient of finding an electron in a subband has been
derived within the effective-mass approximation using the Krieger and
Iafrate equation.  We have clarified the role of the ODEM-induced
inter-subband transitions on the acceleration of the electrons by
defining the effective dispersion.  Using a simplified path
integration method, we have calculated the drift velocity of 2DEG.

One of the important results is that the initial acceleration of an
electron effectively follows the bulk dispersion relation.  
When analyzing systems with a non-zero ODEM using semiclassical approaches -- such as Monte Carlo methods -- care must be taken in the treatment of free-flight processes. In contrast, conventional methods can be safely applied to devices with zero ODEM, such as [100]-channel Si devices on the (001) plane, which are commonly used in industry. For systems with a non-zero ODEM, the applicability of conventional methods can be assessed by considering the critical time $t_\mathrm{c}$. However, it is important to note that $t_\mathrm{c}$ of Eq.~(\ref{eq:2DEG:tc}) derived in this study is based on a two-dimensional system with an infinite quantum well. Therefore, more detailed analyses are required to obtain accurate quantitative values or to address one-dimensional systems.

Combining
the present results with the already known results,\cite{Stern1967}
namely that the density-of-states mass of a two-dimensional system
becomes that of the bulk in the weak confinement limit, it is expected
that the transport properties of the low-dimensional system correctly
approach those of the bulk when many subbands are occupied.  However,
the present study focuses mainly on the dispersion relations, and the
transport calculation is performed only with a simplified path
integration method for 2DEG.  To obtain a definitive answer, it is
necessary to perform detailed transport calculations including
realistic scattering mechanisms as well as the ODEM-induced
inter-subband transitions using an appropriate methodology such as the
Monte Carlo method combined with the KI equation.\cite{Nilsson2001,Hathwar2016,Miyazaki2024,Zhu2024}
This is beyond the scope of the present work and is left for a future
study.

\appendix*
\section{Choice of Gauge}

In the present study, the in-plane electric field is introduced via a
time-dependent vector potential, in accordance with the original
Krieger–Iafrate formulation. In this appendix, we demonstrate that,
for the case of a two-dimensional electron gas (2DEG),
Eq.~(\ref{eq:2DEG:KI}) can also be derived when the in-plane electric
field is introduced through a scalar potential.

We introduce an in-plane electric field via a scalar potential
$e\bm{F}\cdot\bm{x}$. The time-dependent effective mass equation is
then written as
\begin{align}
  i\hbar\frac{\partial}{\partial t}\psi(\bm{x}, z, t)
  =
  [H + e\bm{F}\cdot\bm{x}] \psi(\bm{x}, z, t).
\end{align}
We expand $\psi(\bm{x}, z, t)$ in terms of eigenfuntions
$\phi_{n\bm{k}}(\bm{x}, z)$ of $H$, defined by
Eqs.~(\ref{eq:2DEG:phink}) and (\ref{eq:2DEG:un}), as
\begin{align}
  \psi(\bm{x}, z, t)
  =
  \sum_{n\bm{k}}
  c_{n\bm{k}}(t)
  \phi_{n\bm{k}}(\bm{x}, z).
\end{align}
Substituting into the equation, we obtain
\begin{align}
  &
  i\hbar
  \sum_{n\bm{k}}
  \left[
  \frac{\mathrm{d} c_{n\bm{k}}(t)}{\mathrm{d} t}
  \phi_{n\bm{k}}(\bm{x}, z)
  +
  c_{n\bm{k}}(t)
  \frac{\mathrm{d}\bm{k}}{\mathrm{d}t}
  \cdot
  \frac{\partial \phi_{n\bm{k}}(\bm{x}, z)}{\partial \bm{k}}
  \right]
  \nonumber
  \\
  &
  \qquad
  =
  \sum_{n\bm{k}}
  c_{n\bm{k}}(t)[E_{n\bm{k}} + e \bm{F}\cdot\bm{x}]\phi_{n\bm{k}}(\bm{x}, z).
  \label{eq:Appendix:waveeq0}
\end{align}
From Eqs.~(\ref{eq:2DEG:phink}) and (\ref{eq:2DEG:un}), we derive
the following relation:
\begin{align}
  \frac{\partial \phi_{n\bm{k}}(\bm{x}, z)}{\partial \bm{k}}
  =
  i(\bm{x} - z\bm{\gamma})\phi_{n\bm{k}}(\bm{x}, z).
\end{align}
Substituting this into Eq.~(\ref{eq:Appendix:waveeq0}), we have
\begin{align}
  &
  i\hbar
  \sum_{n\bm{k}}
  \bigg[
  \frac{\mathrm{d} c_{n\bm{k}}(t)}{\mathrm{d} t}
  \phi_{n\bm{k}}(\bm{x}, z)
  \nonumber
  \\
  &
  \qquad
  \qquad
  +
  c_{n\bm{k}}(t)
  \frac{\mathrm{d}\bm{k}}{\mathrm{d}t}
  \cdot
  [i(\bm{x} - z\bm{\gamma})]\phi_{n\bm{k}}(\bm{x}, z)
  \bigg]
  \nonumber
  \\
  &
  \qquad
  =
  \sum_{n\bm{k}}
  c_{n\bm{k}}(t)[E_{n\bm{k}} + e\bm{F}\cdot\bm{x}]\phi_{n\bm{k}}(\bm{x}, z).
\end{align}
Next, we assume the following relation for $\bm{k}$:
\begin{align}
  \hbar
  \frac{\mathrm{d}\bm{k}}{\mathrm{d}t}
  =
  -e\bm{F},
\end{align}
under which the time-dependent effective-mass equation becomes
\begin{align}
  &
  i\hbar
  \sum_{n\bm{k}}
  \left[
  \frac{\mathrm{d} c_{n\bm{k}}(t)}{\mathrm{d} t}
  \phi_{n\bm{k}}(\bm{x}, z)
  -
  \frac{1}{i\hbar}
  c_{n\bm{k}}(t)
  e\bm{F}
  \cdot
  \bm{\gamma}
  z \phi_{n\bm{k}}(\bm{x}, z)
  \right]
  \nonumber
  \\
  &
  \qquad
  =
  \sum_{n\bm{k}}
  c_{n\bm{k}}(t)E_{n\bm{k}}\phi_{n\bm{k}}(\bm{x}, z).
\end{align}
Multiplying both sides from the left by $\phi^*_{n'\bm{k}'}$ and
integrating over all space, we obtain
\begin{align}
  i\hbar
  \frac{\mathrm{d}c_{n\bm{k}}(t)}{\mathrm{d} t}
  -
  \sum_{n'}
  e\bm{F}\cdot\bm{\gamma}
  \langle\zeta_{n}|z|\zeta_{n'}\rangle
  c_{n'\bm{k}}(t)
  =
  E_{n\bm{k}}
  c_{n\bm{k}}(t).
  \label{eq:Appendix:waveeq1}
\end{align}
The result indicates that $c_{n\bm{k}}$ is coupled only to the same
$\bm{k}$.  Therefore, we define a simplified coefficient $c_n$ as
\begin{align}
  c_{n\bm{k}}(t)
  =
  c_{n}(t)
  \exp\left[
  -\frac{i}{\hbar}
  \int_0^t
  E_{n\bm{k}(t')}
  \mathrm{d}t'
  \right].
\end{align}
Substituting this into Eq.~(\ref{eq:Appendix:waveeq1}), we finally
arrive at
\begin{align}
  \frac{\mathrm{d}c_{n}(t)}{\mathrm{d}t}
  =
  -
  i
  \sum_{n'}
  \Omega_{nn'}
  \mathrm{e}^{i\omega_{nn'}t}
  c_{n'}(t).
\end{align}

\begin{acknowledgments}
  The authors thank Prof.\ Laurence Eaves for helpful discussions.
  This work was supported by JSPS KAKENHI Grant Numbers JP20H00250, JP25K01265 and
  JSPS Research Fellow Grant Number JP22J20598.
\end{acknowledgments}

\bigskip

\bibliography{main}

%aipnum4-2.bst 2019-01-14 (MD) hand-edited version of apsrev4-1.bst
%Control: key (0)
%Control: author (8) initials jnrlst
%Control: editor formatted (1) identically to author
%Control: production of article title (0) allowed
%Control: page (1) range
%Control: year (1) truncated
%Control: production of eprint (0) enabled
\providecommand{\noopsort}[1]{}\providecommand{\singleletter}[1]{#1}%
\begin{thebibliography}{35}%
\makeatletter
\providecommand \@ifxundefined [1]{%
 \@ifx{#1\undefined}
}%
\providecommand \@ifnum [1]{%
 \ifnum #1\expandafter \@firstoftwo
 \else \expandafter \@secondoftwo
 \fi
}%
\providecommand \@ifx [1]{%
 \ifx #1\expandafter \@firstoftwo
 \else \expandafter \@secondoftwo
 \fi
}%
\providecommand \natexlab [1]{#1}%
\providecommand \enquote  [1]{``#1''}%
\providecommand \bibnamefont  [1]{#1}%
\providecommand \bibfnamefont [1]{#1}%
\providecommand \citenamefont [1]{#1}%
\providecommand \href@noop [0]{\@secondoftwo}%
\providecommand \href [0]{\begingroup \@sanitize@url \@href}%
\providecommand \@href[1]{\@@startlink{#1}\@@href}%
\providecommand \@@href[1]{\endgroup#1\@@endlink}%
\providecommand \@sanitize@url [0]{\catcode `\\12\catcode `\$12\catcode
  `\&12\catcode `\#12\catcode `\^12\catcode `\_12\catcode `\%12\relax}%
\providecommand \@@startlink[1]{}%
\providecommand \@@endlink[0]{}%
\providecommand \url  [0]{\begingroup\@sanitize@url \@url }%
\providecommand \@url [1]{\endgroup\@href {#1}{\urlprefix }}%
\providecommand \urlprefix  [0]{URL }%
\providecommand \Eprint [0]{\href }%
\providecommand \doibase [0]{https://doi.org/}%
\providecommand \selectlanguage [0]{\@gobble}%
\providecommand \bibinfo  [0]{\@secondoftwo}%
\providecommand \bibfield  [0]{\@secondoftwo}%
\providecommand \translation [1]{[#1]}%
\providecommand \BibitemOpen [0]{}%
\providecommand \bibitemStop [0]{}%
\providecommand \bibitemNoStop [0]{.\EOS\space}%
\providecommand \EOS [0]{\spacefactor3000\relax}%
\providecommand \BibitemShut  [1]{\csname bibitem#1\endcsname}%
\let\auto@bib@innerbib\@empty
%</preamble>
\bibitem [{\citenamefont {Nye}(1985)}]{Nye1985}%
  \BibitemOpen
  \bibfield  {author} {\bibinfo {author} {\bibfnamefont {J.~F.}\ \bibnamefont
  {Nye}},\ }\href@noop {} {\emph {\bibinfo {title} {Physical Properties of
  Crystals}}}\ (\bibinfo  {publisher} {Oxford Science Publications},\ \bibinfo
  {year} {1985})\BibitemShut {NoStop}%
\bibitem [{\citenamefont {Stern}\ and\ \citenamefont
  {Howard}(1967)}]{Stern1967}%
  \BibitemOpen
  \bibfield  {author} {\bibinfo {author} {\bibfnamefont {F.}~\bibnamefont
  {Stern}}\ and\ \bibinfo {author} {\bibfnamefont {W.}~\bibnamefont {Howard}},\
  }\href@noop {} {\bibfield  {journal} {\bibinfo  {journal} {Phys.\ Rev.}\
  }\textbf {\bibinfo {volume} {163}},\ \bibinfo {pages} {816} (\bibinfo {year}
  {1967})}\BibitemShut {NoStop}%
\bibitem [{\citenamefont {Ando}, \citenamefont {Fowler},\ and\ \citenamefont
  {Stern}(1982)}]{Ando1982}%
  \BibitemOpen
  \bibfield  {author} {\bibinfo {author} {\bibfnamefont {T.}~\bibnamefont
  {Ando}}, \bibinfo {author} {\bibfnamefont {A.}~\bibnamefont {Fowler}},\ and\
  \bibinfo {author} {\bibfnamefont {F.}~\bibnamefont {Stern}},\ }\href@noop {}
  {\bibfield  {journal} {\bibinfo  {journal} {Rev. Mod. Phys.}\ }\textbf
  {\bibinfo {volume} {54}},\ \bibinfo {pages} {437} (\bibinfo {year}
  {1982})}\BibitemShut {NoStop}%
\bibitem [{\citenamefont {Esseni}\ \emph {et~al.}(2009)\citenamefont {Esseni},
  \citenamefont {Conzatti}, \citenamefont {De~Michielis}, \citenamefont
  {Serra}, \citenamefont {Palestri},\ and\ \citenamefont
  {Selmi}}]{Esseni2009a}%
  \BibitemOpen
  \bibfield  {author} {\bibinfo {author} {\bibfnamefont {D.}~\bibnamefont
  {Esseni}}, \bibinfo {author} {\bibfnamefont {F.}~\bibnamefont {Conzatti}},
  \bibinfo {author} {\bibfnamefont {M.}~\bibnamefont {De~Michielis}}, \bibinfo
  {author} {\bibfnamefont {N.}~\bibnamefont {Serra}}, \bibinfo {author}
  {\bibfnamefont {P.}~\bibnamefont {Palestri}},\ and\ \bibinfo {author}
  {\bibfnamefont {L.}~\bibnamefont {Selmi}},\ }\href@noop {} {\bibfield
  {journal} {\bibinfo  {journal} {J. Comput. Electron.}\ }\textbf {\bibinfo
  {volume} {8}},\ \bibinfo {pages} {209} (\bibinfo {year} {2009})}\BibitemShut
  {NoStop}%
\bibitem [{\citenamefont {Esseni}, \citenamefont {Palestri},\ and\
  \citenamefont {Selmi}(2011)}]{Esseni2011}%
  \BibitemOpen
  \bibfield  {author} {\bibinfo {author} {\bibfnamefont {D.}~\bibnamefont
  {Esseni}}, \bibinfo {author} {\bibfnamefont {P.}~\bibnamefont {Palestri}},\
  and\ \bibinfo {author} {\bibfnamefont {L.}~\bibnamefont {Selmi}},\
  }\href@noop {} {\emph {\bibinfo {title} {Nnanoscale MOS Transistors}}}\
  (\bibinfo  {publisher} {Cambridge University Press},\ \bibinfo {year}
  {2011})\BibitemShut {NoStop}%
\bibitem [{\citenamefont {Takagi}\ \emph {et~al.}(1994)\citenamefont {Takagi},
  \citenamefont {Toriumi}, \citenamefont {Iwase},\ and\ \citenamefont
  {Tango}}]{Takagi1994}%
  \BibitemOpen
  \bibfield  {author} {\bibinfo {author} {\bibfnamefont {S.}~\bibnamefont
  {Takagi}}, \bibinfo {author} {\bibfnamefont {A.}~\bibnamefont {Toriumi}},
  \bibinfo {author} {\bibfnamefont {M.}~\bibnamefont {Iwase}},\ and\ \bibinfo
  {author} {\bibfnamefont {H.}~\bibnamefont {Tango}},\ }\href@noop {}
  {\bibfield  {journal} {\bibinfo  {journal} {IEEE Trans. Electron Devices}\
  }\textbf {\bibinfo {volume} {41}},\ \bibinfo {pages} {2363} (\bibinfo {year}
  {1994})}\BibitemShut {NoStop}%
\bibitem [{\citenamefont {Laux}(2004)}]{Laux2004}%
  \BibitemOpen
  \bibfield  {author} {\bibinfo {author} {\bibfnamefont {S.}~\bibnamefont
  {Laux}},\ }in\ \href@noop {} {\emph {\bibinfo {booktitle} {IEDM Technical
  Digest.}}}\ (\bibinfo {year} {2004})\ p.\ \bibinfo {pages} {135}\BibitemShut
  {NoStop}%
\bibitem [{\citenamefont {Low}\ \emph {et~al.}(2004)\citenamefont {Low},
  \citenamefont {Li}, \citenamefont {Fan}, \citenamefont {Ng}, \citenamefont
  {Yeo}, \citenamefont {Zhu}, \citenamefont {Chin}, \citenamefont {Chan},\ and\
  \citenamefont {Kwong}}]{Low2004}%
  \BibitemOpen
  \bibfield  {author} {\bibinfo {author} {\bibfnamefont {T.}~\bibnamefont
  {Low}}, \bibinfo {author} {\bibfnamefont {M.}~\bibnamefont {Li}}, \bibinfo
  {author} {\bibfnamefont {W.}~\bibnamefont {Fan}}, \bibinfo {author}
  {\bibfnamefont {S.}~\bibnamefont {Ng}}, \bibinfo {author} {\bibfnamefont
  {Y.-C.}\ \bibnamefont {Yeo}}, \bibinfo {author} {\bibfnamefont
  {C.}~\bibnamefont {Zhu}}, \bibinfo {author} {\bibfnamefont {A.}~\bibnamefont
  {Chin}}, \bibinfo {author} {\bibfnamefont {L.}~\bibnamefont {Chan}},\ and\
  \bibinfo {author} {\bibfnamefont {D.}~\bibnamefont {Kwong}},\ }in\ \href@noop
  {} {\emph {\bibinfo {booktitle} {IEDM Technical Digest.}}}\ (\bibinfo {year}
  {2004})\ p.\ \bibinfo {pages} {151}\BibitemShut {NoStop}%
\bibitem [{\citenamefont {Irie}\ \emph {et~al.}(2004)\citenamefont {Irie},
  \citenamefont {Kita}, \citenamefont {Kyuno},\ and\ \citenamefont
  {Toriumi}}]{Irie2004}%
  \BibitemOpen
  \bibfield  {author} {\bibinfo {author} {\bibfnamefont {H.}~\bibnamefont
  {Irie}}, \bibinfo {author} {\bibfnamefont {K.}~\bibnamefont {Kita}}, \bibinfo
  {author} {\bibfnamefont {K.}~\bibnamefont {Kyuno}},\ and\ \bibinfo {author}
  {\bibfnamefont {A.}~\bibnamefont {Toriumi}},\ }in\ \href@noop {} {\emph
  {\bibinfo {booktitle} {IEDM Technical Digest.}}}\ (\bibinfo {year} {2004})\
  p.\ \bibinfo {pages} {225}\BibitemShut {NoStop}%
\bibitem [{\citenamefont {Tsutsui}\ and\ \citenamefont
  {Hiramoto}(2006)}]{Tsutsui2006}%
  \BibitemOpen
  \bibfield  {author} {\bibinfo {author} {\bibfnamefont {G.}~\bibnamefont
  {Tsutsui}}\ and\ \bibinfo {author} {\bibfnamefont {T.}~\bibnamefont
  {Hiramoto}},\ }\href@noop {} {\bibfield  {journal} {\bibinfo  {journal} {IEEE
  Trans. Electron Devices}\ }\textbf {\bibinfo {volume} {53}},\ \bibinfo
  {pages} {2582} (\bibinfo {year} {2006})}\BibitemShut {NoStop}%
\bibitem [{\citenamefont {Chen}\ \emph {et~al.}(2009)\citenamefont {Chen},
  \citenamefont {Saraya}, \citenamefont {Miyaji}, \citenamefont {Shimizu},\
  and\ \citenamefont {Hiramoto}}]{Chen2009}%
  \BibitemOpen
  \bibfield  {author} {\bibinfo {author} {\bibfnamefont {J.}~\bibnamefont
  {Chen}}, \bibinfo {author} {\bibfnamefont {T.}~\bibnamefont {Saraya}},
  \bibinfo {author} {\bibfnamefont {K.}~\bibnamefont {Miyaji}}, \bibinfo
  {author} {\bibfnamefont {K.}~\bibnamefont {Shimizu}},\ and\ \bibinfo {author}
  {\bibfnamefont {T.}~\bibnamefont {Hiramoto}},\ }\href@noop {} {\bibfield
  {journal} {\bibinfo  {journal} {Jpn. J. Appl. Phys.}\ }\textbf {\bibinfo
  {volume} {48}},\ \bibinfo {pages} {011205} (\bibinfo {year}
  {2009})}\BibitemShut {NoStop}%
\bibitem [{\citenamefont {Mochizuki}\ \emph {et~al.}(2023)\citenamefont
  {Mochizuki}, \citenamefont {Loubet}, \citenamefont {Mirdha}, \citenamefont
  {Durfee}, \citenamefont {Zhou}, \citenamefont {Tsusui}, \citenamefont
  {Frougier}, \citenamefont {Vega}, \citenamefont {Qin}, \citenamefont {Felix},
  \citenamefont {Guo},\ and\ \citenamefont {Bu}}]{Mochizuki2024}%
  \BibitemOpen
  \bibfield  {author} {\bibinfo {author} {\bibfnamefont {S.}~\bibnamefont
  {Mochizuki}}, \bibinfo {author} {\bibfnamefont {N.}~\bibnamefont {Loubet}},
  \bibinfo {author} {\bibfnamefont {P.}~\bibnamefont {Mirdha}}, \bibinfo
  {author} {\bibfnamefont {C.}~\bibnamefont {Durfee}}, \bibinfo {author}
  {\bibfnamefont {H.}~\bibnamefont {Zhou}}, \bibinfo {author} {\bibfnamefont
  {G.}~\bibnamefont {Tsusui}}, \bibinfo {author} {\bibfnamefont
  {J.}~\bibnamefont {Frougier}}, \bibinfo {author} {\bibfnamefont
  {R.}~\bibnamefont {Vega}}, \bibinfo {author} {\bibfnamefont {L.}~\bibnamefont
  {Qin}}, \bibinfo {author} {\bibfnamefont {N.}~\bibnamefont {Felix}}, \bibinfo
  {author} {\bibfnamefont {D.}~\bibnamefont {Guo}},\ and\ \bibinfo {author}
  {\bibfnamefont {H.}~\bibnamefont {Bu}},\ }in\ \href
  {https://doi.org/10.1109/IEDM45741.2023.10413854} {\emph {\bibinfo
  {booktitle} {IEDM Technical Digest.}}}\ (\bibinfo {year} {2023})\ p.~\bibinfo
  {pages} {1}\BibitemShut {NoStop}%
\bibitem [{\citenamefont {Silvestri}\ \emph {et~al.}(2010)\citenamefont
  {Silvestri}, \citenamefont {Reggiani}, \citenamefont {Gnani}, \citenamefont
  {Gnudi},\ and\ \citenamefont {Baccarani}}]{Silvestri2010}%
  \BibitemOpen
  \bibfield  {author} {\bibinfo {author} {\bibfnamefont {L.}~\bibnamefont
  {Silvestri}}, \bibinfo {author} {\bibfnamefont {S.}~\bibnamefont {Reggiani}},
  \bibinfo {author} {\bibfnamefont {E.}~\bibnamefont {Gnani}}, \bibinfo
  {author} {\bibfnamefont {A.}~\bibnamefont {Gnudi}},\ and\ \bibinfo {author}
  {\bibfnamefont {G.}~\bibnamefont {Baccarani}},\ }\href@noop {} {\bibfield
  {journal} {\bibinfo  {journal} {IEEE Trans. Electron Devices}\ }\textbf
  {\bibinfo {volume} {57}},\ \bibinfo {pages} {1567} (\bibinfo {year}
  {2010})}\BibitemShut {NoStop}%
\bibitem [{\citenamefont {Esseni}\ and\ \citenamefont
  {Palestri}(2009)}]{Esseni2009b}%
  \BibitemOpen
  \bibfield  {author} {\bibinfo {author} {\bibfnamefont {D.}~\bibnamefont
  {Esseni}}\ and\ \bibinfo {author} {\bibfnamefont {P.}~\bibnamefont
  {Palestri}},\ }\href@noop {} {\bibfield  {journal} {\bibinfo  {journal} {J.\
  Appl.\ Phys.}\ }\textbf {\bibinfo {volume} {105}},\ \bibinfo {pages} {053702}
  (\bibinfo {year} {2009})}\BibitemShut {NoStop}%
\bibitem [{\citenamefont {Jones}\ and\ \citenamefont
  {Zener}(1934)}]{Jones1934}%
  \BibitemOpen
  \bibfield  {author} {\bibinfo {author} {\bibfnamefont {H.}~\bibnamefont
  {Jones}}\ and\ \bibinfo {author} {\bibfnamefont {C.}~\bibnamefont {Zener}},\
  }\href@noop {} {\bibfield  {journal} {\bibinfo  {journal} {Proc. Roy. Soc.
  London Ser. A}\ }\textbf {\bibinfo {volume} {144}},\ \bibinfo {pages} {0101}
  (\bibinfo {year} {1934})}\BibitemShut {NoStop}%
\bibitem [{\citenamefont {Houston}(1940)}]{Houston1940}%
  \BibitemOpen
  \bibfield  {author} {\bibinfo {author} {\bibfnamefont {W.}~\bibnamefont
  {Houston}},\ }\href@noop {} {\bibfield  {journal} {\bibinfo  {journal} {Phys.
  Rev.}\ }\textbf {\bibinfo {volume} {57}},\ \bibinfo {pages} {184} (\bibinfo
  {year} {1940})}\BibitemShut {NoStop}%
\bibitem [{\citenamefont {Kittel}(1987)}]{Kittel1987}%
  \BibitemOpen
  \bibfield  {author} {\bibinfo {author} {\bibfnamefont {C.}~\bibnamefont
  {Kittel}},\ }\href@noop {} {\emph {\bibinfo {title} {Quantum Theory of
  Solids}}}\ (\bibinfo  {publisher} {John Wiley \& Sons},\ \bibinfo {year}
  {1987})\BibitemShut {NoStop}%
\bibitem [{\citenamefont {Zener}(1934)}]{Zener1934}%
  \BibitemOpen
  \bibfield  {author} {\bibinfo {author} {\bibfnamefont {C.}~\bibnamefont
  {Zener}},\ }\href@noop {} {\bibfield  {journal} {\bibinfo  {journal} {Proc.
  Roy. Soc. A}\ }\textbf {\bibinfo {volume} {145}},\ \bibinfo {pages} {523}
  (\bibinfo {year} {1934})}\BibitemShut {NoStop}%
\bibitem [{\citenamefont {Keldysh}(1958)}]{Keldysh1958}%
  \BibitemOpen
  \bibfield  {author} {\bibinfo {author} {\bibfnamefont {L.}~\bibnamefont
  {Keldysh}},\ }\href@noop {} {\bibfield  {journal} {\bibinfo  {journal} {Sov.
  Phys. JETP}\ }\textbf {\bibinfo {volume} {6}},\ \bibinfo {pages} {763}
  (\bibinfo {year} {1958})}\BibitemShut {NoStop}%
\bibitem [{\citenamefont {Kane}(1959)}]{Kane1959}%
  \BibitemOpen
  \bibfield  {author} {\bibinfo {author} {\bibfnamefont {E.}~\bibnamefont
  {Kane}},\ }\href@noop {} {\bibfield  {journal} {\bibinfo  {journal} {J. Phys.
  Chem. Solids}\ }\textbf {\bibinfo {volume} {12}},\ \bibinfo {pages} {181}
  (\bibinfo {year} {1959})}\BibitemShut {NoStop}%
\bibitem [{\citenamefont {Adams}\ and\ \citenamefont
  {Argyres}(1956)}]{Adams1956}%
  \BibitemOpen
  \bibfield  {author} {\bibinfo {author} {\bibfnamefont {E.}~\bibnamefont
  {Adams}}\ and\ \bibinfo {author} {\bibfnamefont {P.}~\bibnamefont
  {Argyres}},\ }\href@noop {} {\bibfield  {journal} {\bibinfo  {journal} {Phys.
  Rev.}\ }\textbf {\bibinfo {volume} {102}},\ \bibinfo {pages} {605} (\bibinfo
  {year} {1956})}\BibitemShut {NoStop}%
\bibitem [{\citenamefont {Duque-Gomez}\ and\ \citenamefont
  {Sipe}(2012)}]{Duque-Gomez2012}%
  \BibitemOpen
  \bibfield  {author} {\bibinfo {author} {\bibfnamefont {F.}~\bibnamefont
  {Duque-Gomez}}\ and\ \bibinfo {author} {\bibfnamefont {J.~E.}\ \bibnamefont
  {Sipe}},\ }\href@noop {} {\bibfield  {journal} {\bibinfo  {journal} {Phys.
  Rev. A}\ }\textbf {\bibinfo {volume} {85}},\ \bibinfo {pages} {053412}
  (\bibinfo {year} {2012})}\BibitemShut {NoStop}%
\bibitem [{\citenamefont {Fang}, \citenamefont {Duque-Gomez},\ and\
  \citenamefont {Sipe}(2014)}]{Fang2014}%
  \BibitemOpen
  \bibfield  {author} {\bibinfo {author} {\bibfnamefont {Y.}~\bibnamefont
  {Fang}}, \bibinfo {author} {\bibfnamefont {F.}~\bibnamefont {Duque-Gomez}},\
  and\ \bibinfo {author} {\bibfnamefont {J.~E.}\ \bibnamefont {Sipe}},\
  }\href@noop {} {\bibfield  {journal} {\bibinfo  {journal} {Phys. Rev. A}\
  }\textbf {\bibinfo {volume} {90}},\ \bibinfo {pages} {053407} (\bibinfo
  {year} {2014})}\BibitemShut {NoStop}%
\bibitem [{\citenamefont {Chang}\ \emph {et~al.}(2014)\citenamefont {Chang},
  \citenamefont {Potnis}, \citenamefont {Ramos}, \citenamefont {Zhuang},
  \citenamefont {Hallaji}, \citenamefont {Hayat}, \citenamefont {Duque-Gomez},
  \citenamefont {Sipe},\ and\ \citenamefont {Steinberg}}]{Chang2014}%
  \BibitemOpen
  \bibfield  {author} {\bibinfo {author} {\bibfnamefont {R.}~\bibnamefont
  {Chang}}, \bibinfo {author} {\bibfnamefont {S.}~\bibnamefont {Potnis}},
  \bibinfo {author} {\bibfnamefont {R.}~\bibnamefont {Ramos}}, \bibinfo
  {author} {\bibfnamefont {C.}~\bibnamefont {Zhuang}}, \bibinfo {author}
  {\bibfnamefont {M.}~\bibnamefont {Hallaji}}, \bibinfo {author} {\bibfnamefont
  {A.}~\bibnamefont {Hayat}}, \bibinfo {author} {\bibfnamefont
  {F.}~\bibnamefont {Duque-Gomez}}, \bibinfo {author} {\bibfnamefont {J.~E.}\
  \bibnamefont {Sipe}},\ and\ \bibinfo {author} {\bibfnamefont {A.~M.}\
  \bibnamefont {Steinberg}},\ }\href@noop {} {\bibfield  {journal} {\bibinfo
  {journal} {Phys. Rev. Lett.}\ }\textbf {\bibinfo {volume} {112}},\ \bibinfo
  {pages} {170404} (\bibinfo {year} {2014})}\BibitemShut {NoStop}%
\bibitem [{\citenamefont {Krieger}\ and\ \citenamefont
  {Iafrate}(1986)}]{Krieger1986}%
  \BibitemOpen
  \bibfield  {author} {\bibinfo {author} {\bibfnamefont {J.}~\bibnamefont
  {Krieger}}\ and\ \bibinfo {author} {\bibfnamefont {G.}~\bibnamefont
  {Iafrate}},\ }\href@noop {} {\bibfield  {journal} {\bibinfo  {journal} {Phys.
  Rev. B}\ }\textbf {\bibinfo {volume} {33}},\ \bibinfo {pages} {5494}
  (\bibinfo {year} {1986})}\BibitemShut {NoStop}%
\bibitem [{\citenamefont {Bescond}, \citenamefont {Cavassilas},\ and\
  \citenamefont {Lannoo}(2007)}]{Bescond2007}%
  \BibitemOpen
  \bibfield  {author} {\bibinfo {author} {\bibfnamefont {M.}~\bibnamefont
  {Bescond}}, \bibinfo {author} {\bibfnamefont {N.}~\bibnamefont
  {Cavassilas}},\ and\ \bibinfo {author} {\bibfnamefont {M.}~\bibnamefont
  {Lannoo}},\ }\href@noop {} {\bibfield  {journal} {\bibinfo  {journal}
  {Nanotechnol.}\ }\textbf {\bibinfo {volume} {18}},\ \bibinfo {pages} {255201}
  (\bibinfo {year} {2007})}\BibitemShut {NoStop}%
\bibitem [{\citenamefont {Wang}(1999)}]{Wang1999}%
  \BibitemOpen
  \bibfield  {author} {\bibinfo {author} {\bibfnamefont {S.}~\bibnamefont
  {Wang}},\ }\href@noop {} {\bibfield  {journal} {\bibinfo  {journal} {Phys.
  Rev. A}\ }\textbf {\bibinfo {volume} {60}},\ \bibinfo {pages} {262} (\bibinfo
  {year} {1999})}\BibitemShut {NoStop}%
\bibitem [{kc()}]{kc}%
  \BibitemOpen
  \href@noop {} {}\bibinfo {note} {Consider a semiclassical case of an electron
  at rest in the center of a quantum well. When an in-plane electric field
  $\bm{F}$ is applied at $t = 0$, the velocity in the $z$ direction at time $t$
  is given by $v_z(t) = -w_{33} e \bm{F} \cdot \bm{\gamma}t$, so if
  $t_{\mathrm{c}}$ is the time when the electron collides with the quantum-well
  wall, we obtain $t_{\mathrm{c}} = |W_z / w_{33} e \bm{F} \cdot
  \bm{\gamma}t|^{1/2}$. This is equal to $\tau_0$ except for a numerical
  coefficient $\sqrt{\pi}$.}\BibitemShut {Stop}%
\bibitem [{\citenamefont {Chambers}(1952)}]{Chambers1952}%
  \BibitemOpen
  \bibfield  {author} {\bibinfo {author} {\bibfnamefont {R.}~\bibnamefont
  {Chambers}},\ }\href {https://doi.org/10.1088/0370-1298/65/6/114} {\bibfield
  {journal} {\bibinfo  {journal} {Proc. Phys. Soc. (London) A}\ }\textbf
  {\bibinfo {volume} {65}},\ \bibinfo {pages} {458} (\bibinfo {year}
  {1952})}\BibitemShut {NoStop}%
\bibitem [{\citenamefont {Esaki}\ and\ \citenamefont {Tsu}(1970)}]{Esaki1970}%
  \BibitemOpen
  \bibfield  {author} {\bibinfo {author} {\bibfnamefont {L.}~\bibnamefont
  {Esaki}}\ and\ \bibinfo {author} {\bibfnamefont {R.}~\bibnamefont {Tsu}},\
  }\href {https://doi.org/10.1147/rd.141.0061} {\bibfield  {journal} {\bibinfo
  {journal} {IBM J. Res. Develop.}\ }\textbf {\bibinfo {volume} {14}},\
  \bibinfo {pages} {61} (\bibinfo {year} {1970})}\BibitemShut {NoStop}%
\bibitem [{\citenamefont {Iafrate}\ and\ \citenamefont
  {Sokolov}(2020)}]{Iafrate2020}%
  \BibitemOpen
  \bibfield  {author} {\bibinfo {author} {\bibfnamefont {G.}~\bibnamefont
  {Iafrate}}\ and\ \bibinfo {author} {\bibfnamefont {V.}~\bibnamefont
  {Sokolov}},\ }\href {https://doi.org/https://doi.org/10.1002/pssb.201900660}
  {\bibfield  {journal} {\bibinfo  {journal} {phys. stat. solidi (b)}\ }\textbf
  {\bibinfo {volume} {257}},\ \bibinfo {pages} {1900660} (\bibinfo {year}
  {2020})}\BibitemShut {NoStop}%
\bibitem [{\citenamefont {Nilsson}\ \emph {et~al.}(2001)\citenamefont
  {Nilsson}, \citenamefont {Martinez}, \citenamefont {Ghillino}, \citenamefont
  {Sannemo}, \citenamefont {Bellotti},\ and\ \citenamefont
  {Goano}}]{Nilsson2001}%
  \BibitemOpen
  \bibfield  {author} {\bibinfo {author} {\bibfnamefont {H.}~\bibnamefont
  {Nilsson}}, \bibinfo {author} {\bibfnamefont {A.}~\bibnamefont {Martinez}},
  \bibinfo {author} {\bibfnamefont {E.}~\bibnamefont {Ghillino}}, \bibinfo
  {author} {\bibfnamefont {U.}~\bibnamefont {Sannemo}}, \bibinfo {author}
  {\bibfnamefont {E.}~\bibnamefont {Bellotti}},\ and\ \bibinfo {author}
  {\bibfnamefont {M.}~\bibnamefont {Goano}},\ }\href@noop {} {\bibfield
  {journal} {\bibinfo  {journal} {J. Appl. Phys.}\ }\textbf {\bibinfo {volume}
  {90}},\ \bibinfo {pages} {2847} (\bibinfo {year} {2001})}\BibitemShut
  {NoStop}%
\bibitem [{\citenamefont {Hathwar}, \citenamefont {Saraniti},\ and\
  \citenamefont {Goodnick}(2016)}]{Hathwar2016}%
  \BibitemOpen
  \bibfield  {author} {\bibinfo {author} {\bibfnamefont {R.}~\bibnamefont
  {Hathwar}}, \bibinfo {author} {\bibfnamefont {M.}~\bibnamefont {Saraniti}},\
  and\ \bibinfo {author} {\bibfnamefont {S.~M.}\ \bibnamefont {Goodnick}},\
  }\href@noop {} {\bibfield  {journal} {\bibinfo  {journal} {J. Appl. Phys.}\
  }\textbf {\bibinfo {volume} {120}},\ \bibinfo {pages} {044307} (\bibinfo
  {year} {2016})}\BibitemShut {NoStop}%
\bibitem [{\citenamefont {Miyazaki}, \citenamefont {Tanaka},\ and\
  \citenamefont {Mori}(2024)}]{Miyazaki2024}%
  \BibitemOpen
  \bibfield  {author} {\bibinfo {author} {\bibfnamefont {W.}~\bibnamefont
  {Miyazaki}}, \bibinfo {author} {\bibfnamefont {H.}~\bibnamefont {Tanaka}},\
  and\ \bibinfo {author} {\bibfnamefont {N.}~\bibnamefont {Mori}},\ }\href@noop
  {} {\bibfield  {journal} {\bibinfo  {journal} {Jpn. J. Appl. Phys.}\ }\textbf
  {\bibinfo {volume} {63}},\ \bibinfo {pages} {02SP35} (\bibinfo {year}
  {2024})}\BibitemShut {NoStop}%
\bibitem [{\citenamefont {Zhu}\ \emph {et~al.}(2024)\citenamefont {Zhu},
  \citenamefont {Bertazzi}, \citenamefont {Matsubara},\ and\ \citenamefont
  {Bellotti}}]{Zhu2024}%
  \BibitemOpen
  \bibfield  {author} {\bibinfo {author} {\bibfnamefont {M.}~\bibnamefont
  {Zhu}}, \bibinfo {author} {\bibfnamefont {F.}~\bibnamefont {Bertazzi}},
  \bibinfo {author} {\bibfnamefont {M.}~\bibnamefont {Matsubara}},\ and\
  \bibinfo {author} {\bibfnamefont {E.}~\bibnamefont {Bellotti}},\ }\href@noop
  {} {\bibfield  {journal} {\bibinfo  {journal} {J. Appl. Phys.}\ }\textbf
  {\bibinfo {volume} {135}},\ \bibinfo {pages} {065702} (\bibinfo {year}
  {2024})}\BibitemShut {NoStop}%
\end{thebibliography}%

\end{document}